\documentclass[10pt]{article}
\usepackage{latexsym}
\usepackage{amssymb}
\usepackage{amsmath}
\usepackage[dvips]{graphicx}

\title{Recursive approach to supersymmetric quantum mechanics for arbitrary fermion occupation number}
\author{Piotr Korcyl\thanks{e-mail address: korcyl@th.if.uj.edu.pl} \\ \small{\emph{M. Smoluchowski Institute of Physics, Jagiellonian
University}} \\ \small{\emph{Reymonta 4, 30-059 Krak\'{o}w,
Poland}}}
\date{\today}

\begin{document}

\maketitle

\begin{abstract}
We present in details a numerical approach for solving supersymmetric quantum mechanical systems with a gauge symmetry
valid in all fermionic sectors.
The method uses a recursive algorithm to calculate matrix elements of any gauge invariant
operator in the Fock basis, in particular of the Hamiltonian operator,
and can be used for any gauge group. We describe its application
to a supersymmetric anharmonic oscillator model with discrete spectrum.
\end{abstract}

\section{Motivations}
\label{sec. Motivations}

For several years a revival of interest in Supersymmetric Yang-Mills Quantum Mechanics (SYMQM) can be observed.
These supersymmetric quantum mechanical systems
can be obtained, most commonly,
by a dimensional reduction of supersymmetric, $D=d+1$ dimensional,  $\mathcal{N}=1$ Yang-Mills quantum field theories to one point in space.
Such procedure reduces the
local gauge symmetry of the initial field theory to a global symmetry of the reduced quantum mechanical system.
The physical Hilbert space of SYMQM is composed of states invariant under this global symmetry. As well
all pertinent operators must be symmetry singlets. These constraints, being the remnants of the Gauss law, make the analytic construction of
solutions nontrivial.

The growing interest in these systems have several sources.
On one hand, it is due to the their conjectured relation with a particular limit of M-theory \cite{mtheory},
on the other hand, to the regularized dynamics of relativistic quantum membranes and supermembranes which they describe \cite{hoppe}.
Moreover, their bosonic sector can be investigated as a zero-volume
limit of $D=4$, Yang-Mills quantum field theory \cite{luscher1}\cite{luscher2}\cite{baal},
 providing results which can be compared with lattice calculations.
Besides all this, the physically interesting features of SYMQM can be studied on their own.
For these reasons much effort has been devoted to the evaluation of spectra of SYMQM.

Among the variety of approaches, many numerical methods have been adapted to
investigate supersymmetric Yang-Mills quantum mechanics and recently provided new results.
Some of them use path integral picture
of quantum mechanics and Monte Carlo integration \cite{hanada1}\cite{hanada2}, other the Hamiltonian formulation of quantum mechanics and the Fock space
methods \cite{wosiek1}\cite{wosiek2}\cite{wosiek3}. Particularly, basics of the algorithm presented in this paper
were already described in \cite{wosiek4}\cite{wosiek5}. Our approach exploits the Fock space formulation,
and therefore provides a nonperturbative way to calculate the eigenenergies and eigenstates.
It was already applied to many systems uncovering interesting physics \cite{praca_magisterska}.
Although early attempts based on this principle proved to be very helpful
in obtaining qualitative results, their potential was limited due to the rapid growth of the Fock basis.
Comparing to them, the recursive algorithm which we present in this article, not only enables
one to obtain numerical results up to very high precision, and thus, permits more quantitative considerations, but also
makes the calculations in fermionic sectors possible.
The main motivation for it comes from the study of $D=2$, SYMQM.
However, the method is much more general and provides a tool for evaluation of matrix elements of any invariant operator
for any gauge group and in any dimension.
Due to these multiple possibilities of extensions, this paper is the first of a series of articles presenting the results
of studies of the SYMQM systems obtained with our numerical approach. Being the introductory paper,
it contains, apart of the detailed discussion of the recursive algorithm itself, the summary of the
whole framework which will
be needed for the future work. Hence, in the following section we start with an introduction of the basic notions of SYMQM.
Next, we construct the Fock space, which is necessary for
the numerical calculations, and discuss some of its properties. The main part of the paper, the description of the algorithm, is
divided into several sections. First of all, the calculation of the matrix of the scalar products is presented.
Then, the orthonormalization procedure and the automatic evaluation of commutators and anticommutators is described.
Only at that point the formula used for calculation of matrix elements of any operator can be clearly introduced.
Eventually, the full recursive relations will be discussed. We finish by presenting a simple application
of our method to a supersymmetric model with discrete spectrum. Conclusions and an outlook of future research
directions will be provided in the last section.

\section{The framework}
\label{sec. Cut Fock basis framework}

In this section we describe several basic concepts constituting our framework.
We start by introducing quantum mechanics in the cut Fock basis. Then, we define the supersymmetric Yang-Mills quantum
mechanics and derive a particular Hamiltonian which is studied numerically as an illustration of the approach at the end of this article.
In the remaining subsections we discuss the construction of the basis of SYMQM and its properties.

\subsection{Quantum mechanics in a cut Fock basis}

Quantum systems, which are described by a Hamiltonian operator
expressed in terms of position and momentum operators, can be analyzed numerically
in an eigenbasis of occupation number operators - a Fock basis
\cite{wosiek1}. Any occupation number operator can be written as
$a^{\dagger}a$, where $a^{\dagger}$ and $a$ are bosonic
creation and annihilation operators respectively, fulfilling the
well-known commutation relations
\begin{equation}
\label{eq:boz komutatory} [a_p, a_q] = [a_p^{\dagger},
a_q^{\dagger}] = 0,\quad [a_p, a_q^{\dagger}] = \delta_{pq}.
\end{equation}
$p$ and $q$ in \eqref{eq:boz komutatory} are indices which label the
bosonic degrees of freedom.
In the case of supersymmetric system we introduce fermionic creation and
annihilation operators, $f$ and $f^{\dagger}$ respecting the
anticommutation relations
\begin{equation}
\label{eq:ferm komutatory} \{f_p, f_q\} = \{f_p^{\dagger},
f_q^{\dagger}\} = 0 , \quad \{f_p, f_q^{\dagger}\} = \delta_{pq},
\end{equation}
where again $p$ and $q$ are indices which describe the fermionic degrees
of freedom.
Obviously, we also have
\begin{equation}
[f_p, a_q]  =  [f_p^{\dagger}, a_q] = 0, \qquad [f_p, a_q^{\dagger} ] = [f_p^{\dagger}, a_q^{\dagger}] = 0 .
\label{eq:mixed komutatory}
\end{equation}
The momentum and position operators are expressed by creation and
annihilation operators in the usual way
\begin{equation}
\label{eq:op x i y}
x_q = \frac{1}{\sqrt{2}}(a_q+a_q^{\dagger})  ,  \ p_q = \frac{1}{i\sqrt{2}}(a_q-a_q^{\dagger}),
\end{equation}
enabling to express the Hamiltonian operator in terms of bosonic and fermionic creation and annihilation
operators only.

The construction of the basis starts with the
definition of the Fock vacuum $|0\rangle$, as the state fulfilling the conditions
\begin{equation}
a_q |0\rangle = 0, \ f_q |0\rangle = 0, \forall q.
\end{equation}
Any other basis state can be obtained from $|0\rangle$ by a successful action of creation operators.

Eventually, the action of the Hamiltonian operator, which is now an operator function of
\eqref{eq:boz komutatory} and \eqref{eq:ferm komutatory}, is straightforward in such basis.
There is no conceptual difficulties in evaluation of its matrix elements, however such calculations
may turn to be computationally demanding. The recursive algorithm described in this paper
may be a solution to this problem. Once the Hamiltonian matrix is obtained, its eigenvalues correspond simply to the eigenenergies of the quantum
system, and its eigenvectors to the eigenstates.

The numerical analysis requires one last step, namely an introduction of a cut-off $N_{cut}$ on the countably infinite Fock basis,
since it is impossible to deal with infinite matrices on a computer. There are many ways to introduce such a
cut-off depending on the symmetries of the system being investigated.
Finally, we have to perform calculations with several increasing $N_{cut}$
and extract the physical results from the limit of infinite cut-off.
The properties of such a procedure were analyzed in \cite{wosiek1}\cite{maciek+wosiek}\cite{maciek3}\cite{korcyl2}.

\subsection{Supersymmetric Yang-Mills Quantum Mechanics}
\label{sec. Supersymmetric Yang-Mills Quantum Mechanics}


%

The recursive algorithm will be described in the context of a relatively simple, $D=d+1=2$ supersymmetric
Yang-Mills quantum mechanics. In this way we will be able to focus our attention directly on the main features of the algorithm.
However, it can be used for many
different systems and it is conceptually straightforward to generalize it
to the physically more interesting systems like $D=4$ quantum mechanics \cite{wosiek2}\cite{wosiek5}.
In order to obtain the particular Hamiltonian which will be studied in the last section of this article,
we will now derive the general Hamiltonian of $D=2$, SYMQM.
To this end we extend the construction of supersymmetric quantum mechanics \cite{cooper} via
the generalized creation and annihilation operators. We define the latter as
\begin{eqnarray}
A & = & \frac{1}{\sqrt{2}} \big( -i p + W \big), \\
A^{\dagger} & = & \frac{1}{\sqrt{2}} \big( i p + W \big),
\end{eqnarray}
where the superpotential $W=W(x)$ is a function of the position operator $x$. The supercharges are obtained from $A$ and $A^{\dagger}$
as $Q= Af$ and $Q^{\dagger} = A^{\dagger} f^{\dagger}$.
With the idea of SYMQM in mind, a simple generalization is to include a global $SU(N)$ symmetry.
Therefore, we postulate the supercharges to be given by\footnote{Private communication by J. Wosiek}
\begin{equation}
Q  =  \frac{1}{\sqrt{2}} \sum_{a=1}^{N^2-1} \big( -i p_a + W_a \big) f_a,
\label{eq. supercharge 1}
\end{equation}
\begin{equation}
Q^{\dagger} =  \frac{1}{\sqrt{2}} \sum_{a=1}^{N^2-1} \big( i p_a + W_a \big) f^{\dagger}_a,
\label{eq. supercharge 2}
\end{equation}
where $a$ is a color index of the adjoint representation of the $SU(N)$ group, $a=1,\dots,N^2-1$, $f_a, f^{\dagger}_a$ are fermionic operators.
Our system contains now
$N^2-1$ bosonic degrees of freedom, described by $x_a$ and $p_a$, and $N^2-1$ fermionic degrees of freedom, described by $f_a^{\dagger}$ and $f_a$.
The reduced Gauss law restricts the physical Hilbert space to only those states which are invariant under the $SU(N)$ group.
In order to facilitate the construction of such singlets it is useful to introduce a matrix notation. Thanks to the latter all singlets can
be written in terms of traces \cite{maciek1}. Therefore, we define
\begin{eqnarray}
x_{i,j} = \sum_{a=1}^{N^2-1} x_a T^a_{i,j}, \quad && \quad p_{i,j} = \sum_{a=1}^{N^2-1} p_a T^a_{i,j}, \nonumber \\
f^{\dagger}_{i,j} = \sum_{a=1}^{N^2-1} f^{\dagger}_a T^a_{i,j}, \quad && \quad f_{i,j} = \sum_{a=1}^{N^2-1} f_a T^a_{i,j}, \nonumber
\end{eqnarray}
where $T^a_{i,j}$ are the generators of the $SU(N)$ group in the fundamental representation, $i,j = 1, \dots, N$.
Hence all operators
become operator valued matrices.
We also introduce a simplified notation for a trace of any matrix, namely, $\textrm{tr}(O) \equiv (O)$. However, we
will use this notation only when many traces occur and no confusion is induced.

The Hamiltonian of such a system is given by the anticommutator of the supercharges, eqs. \eqref{eq. supercharge 1} and \eqref{eq. supercharge 2},
and its general form reads\footnote{We adopt the notation when a repeated index is assumed to be summed over.}
\begin{equation}
H = \big\{ Q^{\dagger}, Q \big\} = \frac{1}{2} \big( p_a p_a + W_a W_a \big) + \frac{1}{4} \big( \partial_a W_b + \partial_b W_a\big) [f^{\dagger}_a, f_b].
\label{eq. general H}
\end{equation}
In the simplest case, we choose $W_a = x_a$ and obtain a set of $N^2-1$ supersymmetric harmonic oscillators.
For a slightly more complicated case, let us consider
\begin{equation}
W_a=\frac{g}{2} d_{abc} x_b x_c,
\end{equation}
where $g$ is the coupling constant and $d_{abc}$ is the totally symmetric tensor of the $SU(N)$ group.
The Hamiltonian eq.\eqref{eq. general H} reduces to,
\begin{equation}
H = \frac{1}{2} \big( p_a p_a + \frac{g^2}{4} d_{abe} d_{ecd} x_a x_b x_c x_d \big) + \frac{g}{2} d_{abc} x_a [f^{\dagger}_b, f_c].
\label{eq. discrete hamiltonian}
\end{equation}
One of the interesting features of this model is a nontrivial bosonic potential of fourth order which for the $SU(3)$ group is simply
\begin{equation}
 d_{abe} d_{ecd} x_a x_b x_c x_d = \frac{1}{3} \Big( \sum_{a=1}^{8} (x_a)^2 \Big)^2.
\end{equation}
Therefore the system of eq.\eqref{eq. discrete hamiltonian} is expected to have a discrete spectrum of bound states.
In order to rewrite this Hamiltonian in terms of traces we use the following identities valid for any $SU(N)$ group,
\begin{align}
\frac{1}{2}  p_a p_a &= \textrm{tr} \ p^2, \nonumber \\
\frac{1}{8}  x_a x_b x_c x_d d_{abe} d_{ecd} &= \textrm{tr} \ x^4 - \frac{1}{N} (\textrm{tr} \ x^2)^2, \nonumber \\
\frac{1}{2} x_a f^{\dagger}_b f_c d_{abc} &= \textrm{tr}(x f^{\dagger} f) - \textrm{tr}(x f f^{\dagger}), \nonumber
\end{align}
and eventually obtain
\begin{equation}
H = \textrm{tr} \ p^2 + g^2 \Big( \textrm{tr} \ x^4 - \frac{1}{N} (\textrm{tr} \ x^2)^2 \Big) + 2g \ \textrm{tr}(x [ f^{\dagger}, f]).
\end{equation}
We suspend the detailed discussion of this Hamiltonian to section \ref{sec. Applications} where the numerical results
obtained with the recursive algorithm will be presented together with some analytic calculations.
In the following subsection we will describe
the construction of the Fock basis for SYMQM and some of its properties.

\subsection{Elementary bricks and Fock basis of the $SU(N)$ \\ SYMQM}
\label{sec. Elementary bricks and Fock basis of the $SU(N)$ SYMQM}


The Fock states are eigenstates of some occupation number operators. In the case of SYMQM models we consider
gauge-invariant occupation number operators,
\begin{equation}
\textrm{tr} \ a^{\dagger} a = \sum_{q=1}^{N^2-1} a^{\dagger q} a^q,
\label{eq. occupation number operators1}
\end{equation}
\begin{equation}
\textrm{tr} \ f^{\dagger} f = \sum_{q=1}^{N^2-1} f^{\dagger q} f^q.
\label{eq. occupation number operators2}
\end{equation}
Most of the Hamiltonians that we have considered so far conserve the fermionic occupation number.
Therefore, it is physically motivated to construct the Fock basis independently in each subspace
of the physical Hilbert space with a definite fermionic occupation number.
Moreover, states containing
different total number of quanta are orthogonal, so we can further divide the fermionic sectors into subspaces with given number of bosonic
quanta. However, usually the Hamiltonian does not conserve the
bosonic occupation number and thus mixes different bosonic subspaces.
We start the construction of the Fock basis in the purely bosonic situation, and then turn to the fermionic sectors.

\subsubsection{Bosonic elementary bricks}
\label{sec. Bosonic elementary bricks}

A general eigenstate of the occupation number operator, having $n_B$ quanta, can be written as \cite{doktorat_macka}
\begin{equation}
|n_B \rangle \ = \sum_{i_1, \dots, i_{n_B}} T_{i_1, i_2, \dots, i_{n_{B}}} a^{\dagger}_{i_1} a^{\dagger}_{i_2} \dots a^{\dagger}_{i_{n_B}}  |0\rangle,
\end{equation}
where $T$ is a group invariant tensor. It can be shown \cite{maciek1} that
any such invariant tensor can be expressed as linear combination of products of trace
tensors. The latter are just traces of products of $T^a_{i,j}$ matrices,
of which the simplest ones are $\textrm{tr} \ T^a T^b = \frac{1}{2} \delta_{a b}$ and
$\textrm{tr} \ T^a T^b T^c = \frac{1}{4} d_{a b c} + \frac{i}{4} f_{a b c}$.
Therefore,
in matrix notation a state $|n_B\rangle $ can be rewritten as
\begin{equation}
|n_B\rangle \ = \sum_{ \big\{ \sum_{j=2}^N j k_j =n_B \big\} } \gamma_{k_2, \dots, k_{n_B}} ( a^{\dagger 2})^{k_2} ( a^{\dagger 3})^{k_3} \dots ( a^{\dagger n_B})^{k_{n_{B}}}   |0\rangle,
\end{equation}
where $\gamma_{k_2, \dots, k_{n_B}}$ are arbitrary coefficients and the sum is over
all such combinations of exponents $k_j$ that, $\sum_{j=2}^{n_B} j k_j = n_B$,
so that the state $|n_B\rangle$ is composed of $n_B$ quanta.
Notice that even though a single quantum created by $a^{\dagger}_{i,j}$ is gauge dependent,
total numbers of quanta, $n_B$, being the eigenvalues
of the operators eqs.\eqref{eq. occupation number operators1} and \eqref{eq. occupation number operators2},
are gauge independent.

Subsequently, $|n_B\rangle$ can be significantly simplified with the use of
the Cayley-Hamilton theorem.
It states, that any matrix, $A$, satisfies its own characteristic equation.
Therefore, we have for the $SU(2)$, $SU(3)$ and $SU(4)$ groups
\begin{align}
SU(2): &\quad  A^2 -  \frac{1}{2} (A^2) \mathcal{I} = 0, \nonumber \\
\label{eq.cayley-hamilton-relations}
SU(3): &\quad  A^3 - \frac{1}{2} (A^2) A - \frac{1}{3} (A^3) \mathcal{I} = 0, \\
SU(4): &\quad  A^4 - \frac{1}{2} (A^2) A^2 - \frac{1}{3} (A^3) A + \frac{1}{8} (A^2)^2 \mathcal{I} - \frac{1}{4}(A^4) \mathcal{I} = 0. \nonumber
\end{align}
One can use these equalities to reduce traces containing more than $N$ operators of the same kind to simpler ones.
We demonstrate this on an example with $A = a^{\dagger}$ and some arbitrary operator $B$, which can be any operator
involving bosonic or fermionic creation and annihilation operators. Particularly, $B$ can be again a single bosonic creation operator.
Thus, multiplying eqs.\eqref{eq.cayley-hamilton-relations} by $B$ from the right-hand side and taking the trace,
we obtain a set of relations, such as
\begin{eqnarray}
SU(2) && (a^{\dagger}a^{\dagger}B) = \frac{1}{2} (a^{\dagger}a^{\dagger}) (B), \nonumber \\
SU(3) && (a^{\dagger}a^{\dagger}a^{\dagger}B) = \frac{1}{2} (a^{\dagger}a^{\dagger}) (a^{\dagger}B) + \frac{1}{3} (a^{\dagger}a^{\dagger}a^{\dagger}) (B), \\
SU(4) && (a^{\dagger}a^{\dagger}a^{\dagger}a^{\dagger}B) = \frac{1}{2} (a^{\dagger}a^{\dagger}) (a^{\dagger}a^{\dagger}B)
+ \frac{1}{3} (a^{\dagger}a^{\dagger}a^{\dagger}) (a^{\dagger}B)  - \frac{1}{8} (a^{\dagger}a^{\dagger})^2 (B) \nonumber \\
&& \qquad \qquad \qquad +\frac{1}{4} (a^{\dagger}a^{\dagger}a^{\dagger}a^{\dagger}) (B). 
\label{eq. cayley-hamilton-relations-with-b}
\end{eqnarray}
Hence, a general state with $n_B$ quanta for some given $N$, simplifies to
\begin{equation}
|n_B\rangle_N \ = \ \sum_{ \big\{ \sum_{j=2}^N j k_j =n_B \big\} } \gamma_{k_2, \dots, k_N} \ ( a^{\dagger 2})^{k_2} ( a^{\dagger 3})^{k_3} \dots ( a^{\dagger N})^{k_N}|0\rangle,
\label{eq. bosonic state}
\end{equation}
where the traces with more than $N$ creation operators were reduced and the highest trace is now $(a^{\dagger N})$.

We are now in position to define the set of \emph{bosonic elementary bricks}, which is the set of $N-1$ linearly independent
single traces of creation operators, which cannot be further reduced by the Cayley-Hamilton theorem. Table
\ref{tab. elementary_bricks} contains examples of such sets for $N=2$, $N=3$ and $N=4$.
\begin{table}[h!]
\begin{center}
\begin{tabular}{|c|c|c|}
\hline
$SU(2)$ & $SU(3)$ & $SU(4)$ \\
\hline
\hline
$(a^{\dagger}a^{\dagger})$ & $(a^{\dagger}a^{\dagger})$ & $(a^{\dagger}a^{\dagger})$ \\
 & $(a^{\dagger}a^{\dagger}a^{\dagger})$ & $(a^{\dagger}a^{\dagger}a^{\dagger})$ \\
 &  & $(a^{\dagger}a^{\dagger}a^{\dagger}a^{\dagger})$ \\
\hline
\end{tabular}
\caption{Elementary bosonic bricks for $SU(2)$,$SU(3)$ and $SU(4)$. \label{tab. elementary_bricks}}
\end{center}
\end{table}
Products of powers of elementary bosonic bricks acting on the Fock vacuum compose the set of states
\begin{equation}
\Big\{ ( a^{\dagger 2})^{k_2} ( a^{\dagger 3})^{k_3} \dots ( a^{\dagger N})^{k_N}|0\rangle \Big\}_{\sum_{j=2}^{N}j k_j = n_B} \equiv |\big\{n_B\big\}\rangle
\label{eq. bosonic fock basis}
\end{equation}
which spans the subspace of the Hilbert space with $n_B$ bosonic quanta.
We adopted a generalized notation in which $|\big\{n_B\big\}\rangle$ is a vector of all states with $n_B$ quanta.
The
set of states eq.\eqref{eq. bosonic fock basis} is excessively called the Fock basis, though in general it is not orthonormal.
Only after the application of an orthonormalization procedure it will be transformed into a basis.
By acting on it with additional elementary bricks and orthonormalizing, one can obtain the basis in sectors with yet
higher number of bosonic quanta. In this way, starting with the Fock vacuum $|0\rangle$, one can recursively generate the basis for any $n_B$.

\subsubsection{Fermionic bricks}
\label{sec. Fermionic bricks}

The definition of fermionic bricks is a bit more complicated. Besides the fermionic bricks which are single-trace operators,
as, for example, all fermionic operators in the case of the $SU(2)$ symmetry group (see table \ref{tab. su2_elementary_bricks}),
we must also take into account bricks which are multiple-trace operators. They appear in higher fermionic sectors
in the case of symmetry groups with $N>2$ (see tables \ref{tab. su3_elementary_bricks} and \ref{tab. su4_elementary_bricks}).
In order to make this distinction clear we will now introduce some definitions and appropriate notation.

In analogy to the set of elementary bosonic bricks, we define the set of \emph{elementary fermionic bricks}. The latter will contain
all single traces with $n_F$ fermionic creation operators, which cannot be further reduced by the Cayley-Hamilton
theorem\footnote{There is no simple counterpart of the Cayley-Hamilton theorem for anticommuting matrices.
However, an appropriate choice of the $B$ operator in \eqref{eq. cayley-hamilton-relations-with-b} will
produce identities which can be used to simplify or exclude, due to linear dependence, some of the possible composite fermionic bricks \cite{doktorat_macka}.
}. An algorithmic way
to obtain it, is to start with the set of bosonic elementary bricks for a given $N$. Then, for each trace, one should
perform $n_F$ times either one
of the following operations:
replace one bosonic creation operator by a fermionic one or insert one fermionic creation operators into the trace.
In order to obtain a complete basis, the set of elementary bricks must be enlarged by operators,
which are products of fermionic elementary bricks with smaller number of fermionic quanta
and contain $n_F$ fermionic creation operators in total.
In this way we ensure that all possible invariant contractions of $n_F$
fermionic creation operators with a number of bosonic creation operators are considered.
However, the problem of linear dependence appears and one has to pick out only the linearly independent
operators.
The
linear independence can be checked
by an explicit calculation of the determinant of the Gram matrix of states constructed with those bricks.
Fortunately, there exists also an independent, and more direct way of computing the number of Fock basis states
which will be described in subsection \ref{sec. Counting the number of $SU(N)$ SYMQM Fock basis states}.

The enlarged set of all linearly independent fermionic bricks will be refered to as the set of \emph{composite fermionic bricks}.
Its elements will be labeled
by an index $\alpha$, and denoted by $C^{\dagger}(n_B, n_F, \alpha)$, where $n_B$ and $n_F$ are the number of bosonic and fermionic creation
operators, respectively. Such notation is used in order to treat all bosonic and fermionic bricks in the same way.
The additional index $\alpha$ distinguishes the operators with the same $n_F$ and $n_B$.

Moreover, we will denote:
\begin{itemize}
\item the number of composite bricks with $n_F$  fermionic and $n_B$ bosonic quanta by $d(n_F, n_B)$,
\item the total number of composite bricks with $n_F$ fermions by $d(n_F)$.
\end{itemize}
Obviously,
\begin{equation}
d(n_F) = \sum_{n_B} d(n_F, n_B). \nonumber
\end{equation}
We extend our notation to the set of bosonic elementary bricks
\begin{equation}
(a^{\dagger n_B} ) \equiv  C^{\dagger}(n_B, 0). \nonumber
\end{equation}
As an example, Tables \eqref{tab. su2_elementary_bricks},
\eqref{tab. su3_elementary_bricks} and \eqref{tab. su4_elementary_bricks} contain the sets of composite
bricks for $N=2$, $N=3$ and $N=4$ for some fermionic sectors. Let us briefly comment on table \ref{tab. su2_elementary_bricks}.
In this simple case, there is exactly one fermionic operator in each fermionic sector
(see section \ref{sec. Counting the number of $SU(N)$ SYMQM Fock basis states})
and they are just elementary fermionic bricks. One can prove, by simple arguments, that other operators are not possible. For example,
an operator of the form $(f^{\dagger} a^{\dagger} a^{\dagger})$ vanishes identically, since it must involve the totaly antisymmetric tensor
$\epsilon^{i j k}$ summed with a symmetric combination of bosonic operators $a^{\dagger j} a^{\dagger k}$. In analogy, the possible operator
$(f^{\dagger} a^{\dagger})^2$ also vanishes, since it is a square of an anticommuting operator. Such reasonings facilitate the explicit
construction of the set of composite fermionic bricks for gauge groups with $N>2$.
\begin{table}[]
\begin{center}
\begin{tabular}{|c|c|c|}
\hline
$n_F=1$ & $n_F=2$ & $n_F=3$ \\
\hline
\hline
$(f^{\dagger} a^{\dagger})$& $(f^{\dagger}f^{\dagger} a^{\dagger})$ & $(f^{\dagger}f^{\dagger}f^{\dagger})$ \\
\hline
\end{tabular}
\caption{$SU(2)$ fermionic bricks. \label{tab. su2_elementary_bricks}}
\end{center}
\end{table}
\begin{table}
\begin{center}
\begin{tabular}{|c|c|c|c|}
\hline
$n_F=1$ & $n_F=2$ & $n_F=3$ & $n_F=4$\\
\hline
\hline
$(f^{\dagger} a^{\dagger})$& $(f^{\dagger}f^{\dagger} a^{\dagger})$ & $(f^{\dagger}f^{\dagger}f^{\dagger})$ & $(f^{\dagger}f^{\dagger}f^{\dagger}f^{\dagger}a^{\dagger})$\\
$(f^{\dagger}a^{\dagger} a^{\dagger})$ & $(f^{\dagger}f^{\dagger} a^{\dagger}a^{\dagger})$& $(f^{\dagger}f^{\dagger} f^{\dagger}a^{\dagger})$& $(f^{\dagger}a^{\dagger})(f^{\dagger}f^{\dagger}f^{\dagger})$\\
& $(f^{\dagger} a^{\dagger}a^{\dagger}f^{\dagger}a^{\dagger})$& $(f^{\dagger} f^{\dagger}f^{\dagger}a^{\dagger}a^{\dagger})$& $(f^{\dagger}f^{\dagger}f^{\dagger}f^{\dagger}a^{\dagger}a^{\dagger})$\\
& $(f^{\dagger}a^{\dagger})(f^{\dagger} a^{\dagger}a^{\dagger})$& $(f^{\dagger}a^{\dagger})(f^{\dagger} f^{\dagger}a^{\dagger})$&$(f^{\dagger}a^{\dagger}a^{\dagger})(f^{\dagger}f^{\dagger}f^{\dagger})$\\
&& $(f^{\dagger}a^{\dagger}f^{\dagger}f^{\dagger}a^{\dagger}a^{\dagger})$ &$(f^{\dagger}a^{\dagger})(a^{\dagger}f^{\dagger}f^{\dagger}f^{\dagger})$\\
&& $(f^{\dagger}a^{\dagger})(f^{\dagger}f^{\dagger}a^{\dagger}a^{\dagger})$ & $(f^{\dagger}f^{\dagger}a^{\dagger})(f^{\dagger}f^{\dagger}a^{\dagger})$\\
&& $(f^{\dagger}a^{\dagger}a^{\dagger})(f^{\dagger}f^{\dagger}a^{\dagger})$ & $(f^{\dagger}a^{\dagger}a^{\dagger})(f^{\dagger}f^{\dagger}f^{\dagger}a^{\dagger})$\\
&& $(f^{\dagger}a^{\dagger}a^{\dagger})(f^{\dagger}f^{\dagger}a^{\dagger}a^{\dagger})$ & $(f^{\dagger}f^{\dagger}a^{\dagger})(f^{\dagger}f^{\dagger}a^{\dagger}a^{\dagger})$\\
&&& $(f^{\dagger}a^{\dagger})(f^{\dagger}a^{\dagger}a^{\dagger})(f^{\dagger}f^{\dagger}a^{\dagger})$\\
&&& $(f^{\dagger}f^{\dagger}a^{\dagger})(f^{\dagger}a^{\dagger}f^{\dagger}a^{\dagger}a^{\dagger})$\\
\hline
\end{tabular}
\end{center}
\caption{$SU(3)$ fermionic bricks. \label{tab. su3_elementary_bricks}}
\end{table}

Once the set of composite fermionic bricks is constructed, it is easy to write down a general state with $n_B$ bosonic
and $n_F$ fermionic quanta for a given gauge group $SU(N)$. One has to take a linear combination of states obtained by applying one
of the composite fermionic bricks with $n_F$ fermionic quanta to a general bosonic Fock state eq.\eqref{eq. bosonic state}.
Hence,
\begin{align}
& |n_B,n_F \rangle_N \ = \ \sum_{\alpha=1}^{d(n_F)} C^{\dagger}(n, n_F, \alpha) \times  \nonumber \\
&\times \sum_{\big\{\sum_{j=2}^N j k_j = n_B-n\big\}} \gamma_{k_2, \dots, k_N}(\alpha) \ C^{\dagger}(2,0)^{k_2} C^{\dagger}(3,0)^{k_3} \dots C^{\dagger}(N,0)^{k_N}|0\rangle,
\label{eq. fermionic state}
\end{align}
where the coefficients $\gamma_{k_2, \dots, k_N}(\alpha)$ can depend now on $\alpha$.
In order to ensure that the total number of bosonic quanta is $n_B$, we have apply the operator $C^{\dagger}(n, n_F, \alpha)$ containing $n$
bosonic creation operators, to a purely bosonic state with $n_B-n$ quanta.
In analogy to the bosonic case, we can define the set of states,
\begin{align}
\Big\{ C^{\dagger}(n, n_F, \alpha) C^{\dagger}(2,0)^{k_2} \dots C^{\dagger}(N,0)^{k_N} |0\rangle \Big\}_{\sum_{j=2}^{N}j k_j +n = n_B} \equiv |\big\{n_B,n_F\big\}\rangle
\label{eq. fermionic fock basis}
\end{align}
which after orthonormalization will give the basis in the subspace of Hilbert space with $n_B$ and $n_F$ bosonic and fermionic quanta, respectively.
The linear independence and completeness of the set of composite fermionic
bricks ensures that eq.\eqref{eq. fermionic fock basis} form indeed a complete set of states in the fermionic sectors.
\begin{table}[!ht]
\begin{center}
\begin{tabular}{|c|c|c|}
\hline
$n_F=1$ & $n_F=2$ \\
\hline
\hline
$(f^{\dagger} a^{\dagger})$& $(f^{\dagger}f^{\dagger}a^{\dagger})$ \\
$(f^{\dagger}a^{\dagger} a^{\dagger})$ & $(f^{\dagger}f^{\dagger}a^{\dagger}a^{\dagger})$ \\
$(f^{\dagger} a^{\dagger}a^{\dagger}a^{\dagger})$& $(f^{\dagger}f^{\dagger}a^{\dagger}a^{\dagger}a^{\dagger})$ \\
 & $(f^{\dagger}a^{\dagger}f^{\dagger}a^{\dagger}a^{\dagger})$ \\
 & $(f^{\dagger}a^{\dagger})(f^{\dagger}a^{\dagger}a^{\dagger})$ \\
 & $(f^{\dagger}a^{\dagger}f^{\dagger}a^{\dagger}a^{\dagger}a^{\dagger})$ \\
 & $(f^{\dagger}a^{\dagger})(f^{\dagger}a^{\dagger}a^{\dagger}a^{\dagger})$ \\
 & $(f^{\dagger}a^{\dagger}a^{\dagger}f^{\dagger}a^{\dagger}a^{\dagger}a^{\dagger})$  \\
 & $(f^{\dagger}a^{\dagger}a^{\dagger})(f^{\dagger}a^{\dagger}a^{\dagger}a^{\dagger})$ \\
\hline
\end{tabular}
\end{center}
\caption{$SU(4)$ fermionic bricks. \label{tab. su4_elementary_bricks}}
\end{table}

\subsection{Counting the number of $SU(N)$ SYMQM Fock basis states}
\label{sec. Counting the number of $SU(N)$ SYMQM Fock basis states}

In the preceding section we have shown, that the Fock basis in any sector is given
by eq.\eqref{eq. bosonic fock basis} or eq.\eqref{eq. fermionic fock basis}.
As an example, table \ref{tab. su2 basis} contains few simplest states of
the Fock basis with the $SU(2)$ symmetry.
\begin{table}[h!]
\begin{center}
\begin{tabular}{|c||c|c|c|c|}
\hline
$n_B$ & $n_F=0$ & $n_F=1$ & $n_F=2$ & $n_F=3$\\
\hline
\hline
0 & $|0\rangle$ & & & $(f^{\dagger}f^{\dagger}f^{\dagger})|0\rangle$\\
\hline
1 & & $(f^{\dagger} a^{\dagger})|0\rangle$ & $(f^{\dagger} f^{\dagger} a^{\dagger})|0\rangle$ & \\
\hline
2 & $|2\rangle \equiv (a^{\dagger} a^{\dagger})|0\rangle$ & & & $(f^{\dagger} f^{\dagger} f^{\dagger})|2\rangle$ \\
\hline
3 & & $(f^{\dagger} a^{\dagger}) |2\rangle$ & $(f^{\dagger} f^{\dagger} a^{\dagger})|2\rangle$ & \\
\hline
\vdots & & & & \\
\hline
2n & $|2n\rangle \equiv (a^{\dagger} a^{\dagger})^n|0\rangle$ & & & $(f^{\dagger} f^{\dagger} f^{\dagger})|2n\rangle$ \\
\hline
2n+1 & & $(f^{\dagger} a^{\dagger}) |2n\rangle$ & $(f^{\dagger} f^{\dagger} a^{\dagger})|2n\rangle$ & \\
\hline
\vdots &&&&\\
\hline
\end{tabular}
\caption{Construction of the basis for the $SU(2)$ gauge group. \label{tab. su2 basis}}
\end{center}
\end{table}

Before considering the set of vectors as in table \eqref{tab. su2 basis}
as a basis one must check its linear independence.
This can be done by explicit calculations of the determinant
of the Gram matrix however such computations become cumbersome for larger number of states.
Fortunately there exist an alternative, group-theoretical way of calculating the total number of
linearly independent gauge-invariant Fock states. It
was suggested by Janik and elaborated by Trzetrzelewski \cite{doktorat_macka}\cite{maciek2}.
Instead of an explicit construction of basis states, this approach exploits the orthogonality of the characters.
Such an alternative method is of great practical value since it may serve as a crosscheck to our recursive algorithm.

\subsubsection{Character method}
\label{sec. Character method}

Let $D(n_B, n_F)$ be the number of gauge-invariant, linearly independent states
with $n_B$ bosonic and $n_F$ fermionic quanta. $D(n_B, n_F)$ can be obtained from the orthogonality relation
of the characters of the $SU(N)$ group.

Each bosonic and fermionic creation operator transforms
according to the adjoint representation
of the $SU(N)$ group. Hence, the products of creation operators,
which are needed for the construction of basis states, transform as products of the adjoint representations.
From the representation theory it is known, that the square of any irreducible representation is reducible and
can be expressed as a sum of a symmetric and antisymmetric parts. This statement written in terms of characters reads,
\begin{eqnarray}
\chi(R) \times \chi(R) &=& [\chi(R) \times \chi(R)] + \{\chi(R) \times \chi(R) \} \nonumber \\
&=& \frac{1}{2}\big( \chi^2(R) + \chi(R^2) \big) + \frac{1}{2}\big( \chi^2(R) - \chi(R^2)\big),
\label{eq. charaktery}
\end{eqnarray}
where the symbols $[ \dots ]$ and $\{ \dots \}$ denote the symmetric and antisymmetric part, respectively,
and $\chi(R^2)$ is the trace of the matrix of the representation $R$ squared.
A generalization of the eq.\eqref{eq. charaktery}
is known as the Fr\"obenius theorem \cite{hamermesh}, and has a practical meaning, since
the characters of powers of $R$ are often explicitly given. It gives the expressions for the symmetrized and antisymmetrized
characters of a product of $p$ representations $R$,
\begin{eqnarray}
[ \times_{k=1}^p \chi(R)] &=& \sum_{\sum_{k=2}^p k i_k = p} \prod_{k=1}^{p} \frac{1}{i_k!} \frac{\chi^{i_k}(R^k)}{k^{i_k}}, \\
\{ \times_{k=1}^p \chi(R) \} &=& \sum_{\sum_{k=2}^p k i_k = p} (-1)^{\sum_{k=2}^p i_k} \prod_{k=1}^{p} \frac{1}{i_k!} \frac{\chi^{i_k}(R^k)}{k^{i_k}},
\end{eqnarray}
where the sum is over all partitions of the number $p$ into numbers $2, \dots, p$, $i_j$ being the multiplicity
of the $j$ number in a given partition.
Thus, the most general product of $n_B$ bosonic and $n_F$ fermionic creation operators will be in the representation, which character is equal to
$[ \times_{k=1}^{n_B} \chi(R)]\{ \times_{k=1}^{n_F} \chi(R)\}$, where $R$ denotes now the adjoint representation of the $SU(N)$ group.
From the orthogonality property of the characters we have
\begin{equation}
D(n_B, n_F) = \int d_{\mu_{SU(N)}} \ 1 \ [ \times_{k=1}^{n_B} \chi(R)]\{ \times_{k=1}^{n_F} \chi(R)\} ,
\label{eq. d}
\end{equation}
where $1$ stands for the character of the trivial representation and $d \mu_{SU(N)}$ is the group invariant measure on $SU(N)$.

A convenient parametrization of the group manifold is by $N^2-1$ Euler angles $\alpha_i$, all defined on $[0, 2 \pi]$.
For example the group elements of $SU(3)$ read \cite{byrd}
\begin{equation}
U = e^{i \lambda_3 \alpha_1}e^{i \lambda_2 \alpha_2}e^{i \lambda_3 \alpha_3}e^{i \lambda_5 \alpha_4}e^{i \lambda_3 \alpha_5}e^{i \lambda_2 \alpha_6}e^{i \lambda_3 \alpha_7}e^{i \lambda_8 \alpha_8}
\end{equation}
and the generalization of the above equation to $SU(N)$ can be found in \cite{tilma}.
The last element needed to calculate $D(n_B, n_F)$ are the characters $\chi(R^k)$. They are given by the Weyl formula \cite{weyl}
\begin{equation}
\chi(R) = \sum_{i,j=1}^N e^{i(\alpha_i - \alpha_j)} -1, \qquad \chi(R^k) = \sum_{i,j=1}^N e^{i k(\alpha_i - \alpha_j)} - 1.
\end{equation}
The invariant measure reads \cite{doktorat_macka},
\begin{equation}
d_{\mu_{SU(N)}} = \frac{1}{N!} \prod_{i=1}^N \frac{d\alpha_i}{2 \pi} \big| \prod_{i<j} (e^{i\alpha_i} - e^{i\alpha_j}) \big|^2 \delta_P \big( \sum_{i=1}^N \alpha_i \big),
\end{equation}
where $\delta_P(x)$ is a periodic delta-function given by
\begin{equation}
\delta_P \big(\sum_{i=1}^N \alpha_i \big) = \sum_{k=-\infty}^{\infty} \delta(\sum_{i=1}^N \alpha_i - 2 \pi k),
\end{equation}
with $k$ integer.

\subsubsection{Generating functions for $D(n_B, n_F)$}
\label{sec. Generating functions for $D(n_B, n_F)$}

Eq. \eqref{eq. d} is difficult to evaluate for any $N$, however it was calculated for few simplest groups \cite{doktorat_macka}.
In these cases, the numbers $D(n_B, n_F)$ can be encoded in a generating function with two
parameters $t$ and $s$, $G(t,s)$,
\begin{equation}
G(t,s) = \sum_{n_B,n_F} D(n_B,n_F) t^{n_B} (-s)^{n_F},
\end{equation}
which is very useful in practical applications.
For $N=3$,
$G(t,s)$ can be expressed in terms of simple
polynomials in $t$ \cite{doktorat_macka}, namely
\begin{equation}
G(t,s) = \Big( \prod_{k=2}^N \frac{1}{1-t^k}\Big) \sum_{i=0}^{N^2-1} (-1)^i s^i c_i(t),
\end{equation}
and the polynomials $c_i(t)$ read,
\begin{eqnarray}
c_0(t) &=& 1,\nonumber \\
c_1(t) &=& t + t^2,\nonumber \\
c_2(t) &=& t + t^2 + 2t^3,\nonumber \\
c_3(t) &=& 1+t+2t^2+3t^3+t^4,\nonumber \\
c_4(t) &=& 2t + 4t^2 + 2t^3 + 2t^4,\nonumber \\
c_{8-i}(t) &=& c_i(t).
\label{eq. su3 polynomials}
\end{eqnarray}
In this form some information contained in $G(t,s)$ become evident.
The term proportional to $s^0$ is equal to the generating function for the number
of partitions into numbers $\big\{N, N-1, \dots, 2\big\}$. Obviously, there are as many states with $n_B$ quanta
as there are ways of obtaining $n_B$ from multiples
of the numbers of quanta contained in the elementary bosonic bricks.
Furthermore, from the polynomials
$c_i(t)$ the combinatorial interpretation of the multiplicities of states in the fermionic sectors can be read off.
Particularly, the number $d(n_F, n_B)$ of composite fermionic bricks with $n_B$
bosonic quanta in a given fermionic sector is simply given by,
\begin{equation}
d(n_F, n_B) = \frac{1}{n_B!}\frac{d^{n_B}}{dt^{n_B}} c_{n_F}(t)\Bigg|_{t=0}.
\end{equation}
As an example, let us take the polynomial $c_1(t)$ for the $SU(3)$ group. We have one brick
with a single bosonic quantum - the $(a^{\dagger}f^{\dagger})$ brick, and one brick with
two bosonic quanta - the $(a^{\dagger}a^{\dagger}f^{\dagger})$ brick. A less trivial example
is given by the $c_2(t)$ polynomial. Apart of the two operators with a single and double
bosonic quanta, $(a^{\dagger} f^{\dagger} f^{\dagger})$ and $(a^{\dagger}a^{\dagger}f^{\dagger} f^{\dagger})$, respectively,
we now have two operators with three bosonic quanta, namely,
$(f^{\dagger}a^{\dagger}a^{\dagger}f^{\dagger}a^{\dagger})$ and $(f^{\dagger}a^{\dagger})(f^{\dagger}a^{\dagger}a^{\dagger})$.


Thus, we can crosscheck the number of basis states obtained by direct construction and elimination of linearly dependent states
with the one computed using the above group-theoretical predictions. Table \ref{tab. su3 basis}
 presents the multiplicity of basis states with
given $n_B$ and $n_F$ quanta for $N=3$ up to $n_B=30$, calculated with both methods.
We simply give a single set of numbers since the results agree exactly.
\begin{table}[h!]
\begin{center}
\begin{tabular}{|c||c|c|c|c|c|c|c|c|c|}
\hline
$n_B$ & $0$ & $1$ & $2$ & $3$ & $4$ & $5$ & $6$ & $7$ & $8$ \\
\hline
0 & 1 & & & 1 & & 1 & & & 1\\
1 &  & 1 & 1 & 1 & 2 & 1 & 1 & 1 & \\
2 & 1 & 1 & 1 & 3 & 4 & 3 & 1 & 1 & 1 \\
3 & 1 & 1 & 3 & 5 & 4 & 5 & 3 & 1 & 1 \\
4 & 1 & 2 & 2 & 5 & 8 & 5 & 2 & 2 & 1 \\
5 & 1 & 2 & 4 & 7 & 8 & 7 & 4 & 2 & 1 \\
6 & 2 & 2 & 4 & 9 & 10 & 9 & 4 & 2 & 2 \\
7 & 1 & 3 & 5 & 9 & 12 & 9 & 5 & 3 & 1 \\
8 & 2 & 3 & 5 & 11 & 14 & 11 & 5 & 3 & 2 \\
9 & 2 & 3 & 7 & 13 & 14 & 13 & 7 & 3 & 2 \\
10 & 2 & 4 & 6 & 13 & 18 & 13 & 6 & 4 & 2 \\
11 & 2 & 4 & 8 & 15 & 18 & 15 & 8 & 4 & 2\\
12 & 3 & 4 & 8 & 17 & 20 & 17 & 8 & 4 & 3 \\
13 & 2 & 5 & 9 & 17 & 22 & 17 & 9 & 5 & 2 \\
14 & 3 & 5 & 9 & 19 & 24 & 19 & 9 & 5 & 3 \\
15 & 3 & 5 & 11 & 21 & 24 & 21 & 11 & 5 & 3 \\
16 & 3 & 6 & 10 & 21 & 28 & 21 & 10 & 6 & 3 \\
17 & 3 & 6 & 12 & 23 & 28 & 23 & 12 & 6 & 3 \\
18 & 4 & 6 & 12 & 25 & 30 & 25 & 12 & 6 & 4 \\
19 & 3 & 7 & 13 & 25 & 32 & 25 & 13 & 7 & 3 \\
20 & 4 & 7 & 13 & 27 & 34 & 27 & 13 & 7 & 4 \\
21 & 4 & 7 & 15 & 29 & 34 & 29 & 15 & 7 & 4\\
22 & 4 & 8 & 14 & 29 & 38 & 29 & 14 & 8 & 4 \\
23 & 4 & 8 & 16 & 31 & 38 & 31 & 16 & 8 & 4 \\
24 & 5 & 8 & 16 & 33 & 40 & 33 & 16 & 8 & 5 \\
25 & 4 & 9 & 17 & 33 & 42 & 33 & 17 & 9 & 4 \\
26 & 5 & 9 & 17 & 35 & 44 & 35 & 17 & 9 & 5 \\
27 & 5 & 9 & 19 & 37 & 44 & 37 & 19 & 9 & 5 \\
28 & 5 & 10 & 18 & 37 & 48 & 37 & 18 & 10 & 5 \\
29 & 5 & 10 & 20 & 39 & 48 & 39 & 20 & 10 & 5 \\
30 & 6 & 10 & 20 & 41 & 50 & 41 & 20 & 10 & 6 \\
\hline
\end{tabular}
\caption{Multiplicity of Fock basis states with given number of bosonic and
fermionic quanta for the $SU(3)$ gauge group. These numbers can be obtained from
the generating function eq. \eqref{eq. su3 polynomials} and from the orthonormalization
procedure of the algorithm independently.\label{tab. su3 basis}}
\end{center}
\end{table}

\subsection{Symmetries of the $SU(N)$ SYMQM Fock basis}
\label{sec. Symmetries of the $SU(N)$ SYMQM Fock basis}

The Fock basis of SYMQM, constructed in the way described above, has several nontrivial symmetries which can be interpreted as
announcements of dynamical symmetries of the Hamiltonians of SYMQM.
The most important of them are the supersymmetry and particle-hole symmetry. The former can be observed
as a matching of eigenenergies from neighboring fermionic sectors, while the latter is defined
as a matching of spectra from the sector with $p$ fermions and the sector with $N^2-1-p$ fermions with $0\le p \le N^2-1$.
Let us now describe three observations \cite{doktorat_macka} of nontrivial relations
among the multiplicities $D(n_B, n_F)$ and their interpretations in terms
of these symmetries.

\begin{itemize}
\item \underline{Supersymmetry}

\begin{itemize}
\item
For each $n_B$ there are as many bosonic basis states (states with $n_F$ even) as fermionic basis states (states with $n_F$ odd).
\begin{equation}
\forall_{n_B} \qquad \sum_{n_F \textrm{ - even}} D(n_B, n_F) = \sum_{n_F \textrm{ - odd}} D(n_B, n_F). \nonumber
\end{equation}
Its validity can be checked explicitly for $SU(3)$ case, either in table \ref{tab. su3 basis} by summing
the numbers of states with $n_F$ even and odd in each row separately, or in eqs. \eqref{eq. su3 polynomials} by summing
appropriate polynomials.
This relation can be also exactly proved for any $N$ using the general form of the generating function \cite{doktorat_macka}.

\item
Summing the number of states along
diagonal lines with $n_B + n_F$ or $n_B - n_F$ fixed, for $n_F$ even and $n_F$ odd separately yields the same results,
\begin{align}
\forall_{n_B} \qquad \sum_{n_F \textrm{ - even}} D(n_B \pm n_F, n_F) = \sum_{n_F \textrm{ - odd}} D(n_B \pm n_F, n_F), \nonumber
\end{align}
Such diagonal lines correspond to the action of the supersymmetric gauged harmonic oscillator supercharges, $Q=(fa)$, and $Q^{\dagger}=(f^{\dagger}a^{\dagger})$. They are of special interest since
introducing the cut-offs in the consecutive fermionic sectors according to one of these lines allows to obtain an
exact supersymmetric degeneracy for finite
cut-off.
Again, this result can be proved for any $N$ using the generating functions \cite{doktorat_macka}.
\end{itemize}

This is still not \emph{true} supersymmetry. There is no dynamical supermultiplets, because at this stage we have not defined any Hamiltonian.
However, it is interesting that already at this level we have such a matchings.

\item \underline{Particle-hole symmetry}

\begin{itemize}
\item
For any $n_B$, the number of states in the sector with $n_F$ fermions is equal to the number of states in the sector with $N^2-1-n_F$ fermions,
\begin{equation}
\forall_{n_B} \qquad D(n_B, n_F) = D(n_B , N^2 - 1 - n_F), \qquad n_F = 0, \dots, N^2 - 1. \nonumber
\end{equation}
If we expect that the spectrum in the sector with $n_F$ fermionic quanta coincides with the spectrum in the sector with
$N^2-1-n_F$ then the equality of the multiplicity of basis states in those sectors can be interpreted as a nontrivial announcing of
the particle-hole symmetry.
\end{itemize}

\end{itemize}

Summarizing, already at kinematic level one can find symptoms of the symmetries of the SYMQM systems.

\section{Description of the algorithm}
\label{sec.Description of the algorithm}

The main idea of the recursive algorithm has been already described in \cite{wosiek4}\cite{wosiek5}.
It relies on the observation
that the most efficient way to evaluate a matrix element of an operator is to relate it
to simpler matrix elements of some operators, which have been already evaluated at an earlier stage of calculations.
In this way the explicit construction of
the Fock basis vectors is not necessary.
As an input the algorithm needs
the commutators/anticommutators of elementary bricks and any other invariant operators, which appear in these resulting commutators/anticommutators.

In order to expose the algorithm in a clear way, we will start by explaining the
construction of the Fock basis and then the calculation of the matrix of scalar products. Having such a matrix,
one can orthonormalize the basis vectors. The procedure used to this end will be described in the subsequent subsections.
At that point the formula used
for calculation of matrix elements of any operator will become evident.
Eventually, the full recursive relations will be presented.

\subsection{Recursive construction of the Fock basis}
\label{sec. Recursive construction of the Fock basis}

Although the Fock states are not explicitly needed for the computation of matrix elements
of an operator, they do appear in the labeling of those matrix elements. Hence, we
should have a recursive way of obtaining the Fock basis. Let us assume that such basis
is already constructed in the sectors with the number of bosonic and
fermionic quanta smaller than $n_B$ and $n_F$ respectively. Then, the Fock basis in the sector with
$n_B$ bosonic and $n_F$ fermionic quanta can be created as the sum of all states obtained
by the action of appropriate bricks on the already generated Fock basis states. In our generalized notation this can be written as
\begin{equation}
|\big\{n_B,n_F\big\}\rangle = \sum_{k=2}^{N} C^{\dagger}(k,0) |\big\{n_B-k,n_F\big\}\rangle. \nonumber
\end{equation}
Note that in general such states
will not form an orthonormal set of states. Moreover, the same state may appear in several copies,
differing in the order of successive bricks used to build it. Those duplicates will be treated as distinct states.
The basis is obtained once this redundancy is removed and the remaining states orthonormalized.

\subsection{Matrix of scalar products}
\label{sec. Matrix of scalar products}

The Fock basis obtained recursively form a complete set of states which are not orthonormalized. Thus, one has to calculate the matrix of scalar products.
It is sufficient to calculate the scalar products among the states containing
a given number of quanta, $n_B$ and $n_F$, since those having different number of bosonic or fermionic quanta
are orthogonal by definition.
The matrix of such scalar products will be denoted by $S(n_{B}, n_{F})$.\\

The definition of $S(n_B, n_F)$ differs from the standard definition of the Gram matrix in few aspects.
The traditional Gram matrix contains the scalar products of linearly
independent states and for our systems it is a $D(n_B, n_F) \times D(n_B, n_F)$ matrix.
Contrary, since in our algorithm the Fock basis is defined recursively,
the $S(n_B, n_F)$ is the matrix of scalar products of states that have $n_B$ bosonic and $n_F$ fermionic
quanta and are obtained by the action of appropriate bricks on
states from sectors with smaller number of bosonic and fermionic quanta.
Therefore, the matrix $S(n_B, n_F)$ is usually bigger
than the Gram matrix since some of the states can be included several times. Such redundancy is best illustrated by the following example.
For the
$SU(3)$ gauge group, there is only one Fock state with $5$ bosonic quanta, $|\big\{5,0\big\}\rangle = C^{\dagger}(2,0)C^{\dagger}(3,0) |0\rangle$. However,
the calculation of the matrix of scalar products $S(5,0)$ yields a $2 \times 2$ matrix instead of a single number, namely,
\begin{equation}
S(5,0) = \langle  \big\{5,0\big\} | \big\{5,0\big\} \rangle \longrightarrow \left( \begin{array}{cc}
S(5,0)_{3,3} & S(5,0)_{3,2}\\
S(5,0)_{2,3} & S(5,0)_{2,2}
\end{array} \right) , \nonumber
\label{eq. przyklad 2}
\end{equation}
where the different matrix elements $S(5,0)_{i,j}$ correspond to the multiple possibilities of pulling out an elementary brick out of the state $|\big\{5,0\big\}\rangle$,
\begin{equation}
|\big\{5,0\big\} \rangle = C^{\dagger}(2,0) |\alpha \rangle = C^{\dagger}(3,0) |\beta \rangle, \nonumber
\end{equation}
where $|\alpha \rangle$ and $|\beta \rangle$ are appropriate remaining states.
Such doubling must be eliminated. The procedure which we used to achieve this will be described in the next subsection.\\


The evaluation of $S(n_B, n_F)$ can be divided into two separate cases.\\

If $n_F = 0$, we pull out one bosonic elementary brick from each of the basis states. Since we have $N-1$ bosonic elementary bricks
(see tables \ref{tab. su2_elementary_bricks}, \ref{tab. su3_elementary_bricks}, \ref{tab. su4_elementary_bricks}),
there is $N-1$ different ways to do this, if only $n_B \ge N$. If $n_B < N$ we can pull out
only $n_B-1$ different bosonic elementary bricks.
Thus, in a generic situation, $S(n_{B},0)$ will
be a $(N-1) \times (N-1)$ matrix of the form,
\begin{align}
S(n_B,0)= \left( \begin{array}{cccc}
 S(n_B,0)_{2,2}  & S(n_B,0)_{2,3} & \dots & S(n_B,0)_{2,N-1}\\
 S(n_B,0)_{3,2}  & S(n_B,0)_{3,3} & \dots & S(n_B,0)_{3,N-1}\\
& & \vdots & \\
 S(n_B,0)_{N-1,2} & S(n_B,0)_{N-1,3} & \dots & S(n_B,0)_{N-1,N-1}
\end{array} \right) \nonumber
\end{align}
where
\begin{equation}
S(n_B,0)_{p,q} \equiv \langle \big\{n_B-p,0\big\}| C(p,0) C^{\dagger}(q,0) | \big\{n_B-q,0\big\} \rangle. \nonumber
\end{equation}
The matrix element, calculated by extracting the
$C(p,0)$ elementary brick from the left state and the $C^{\dagger}(q,0)$ elementary brick from the right state,
denoted by $S(n_B,0)_{p,q}$, can be expressed in terms of matrix elements
of operators between basis states with lower number of bosonic quanta as,
\begin{eqnarray}
S(n_B,0)_{p,q}  \
&=& \ \langle \big\{n_{B} - p,0\big\}| \big[ C(p,0), C^{\dagger}(q,0) \big] |\big\{n_{B} - q,0\big\} \rangle \nonumber \\
&+& \langle \big\{n_{B} - p,0\big\}| C^{\dagger}(q,0) C(p,0) | \big\{n_{B} - q,0\big\}\rangle .
\label{eq.scalar product bosonic relation}
\end{eqnarray}
Hence, we expressed $S$ from the sector with $n_B$ quanta
in terms of matrix elements of operators evaluated in the sector with $n_B - q$ quanta.
The procedure to calculate these matrix elements is described in subsection \ref{sec. Matrix elements of gauge invariant operators}.\\

If $n_F \ne 0$, we first want to anticommute the composite fermionic bricks with $n_F$ fermionic quanta.
Again, in a generic case, we treat $S(n_{B}, n_{F})$  as a matrix of size $d(n_F) \times d(n_F)$ and of the form
\begin{align}
S(n_B,n_F) &= \nonumber \\
&\left( \begin{array}{cccc}
S(n_B,n_F)_{1, 1} & S(n_B,n_F)_{1,2} & \dots & S(n_B,n_F)_{1,d(n_F)}\\
S(n_B,n_F)_{2, 1} & S(n_B,n_F)_{2,2} & \dots & S(n_B,n_F)_{2,d(n_F)}\\
& & \vdots & \\
S(n_B,n_F)_{d(n_F),1} & S(n_B,n_F)_{d(n_F),2} & \dots & S(n_B,n_F)_{d(n_F),d(n_F)}
\end{array} \right) \nonumber
\end{align}
where
\begin{equation}
S(n_B,n_F)_{p,q} \equiv \langle \big\{n_B-n_p,0\big\}| C(n_p,n_F,p) C^{\dagger}(n_q,n_F,q) | \big\{n_B-n_q,0\big\} \rangle. \nonumber
\end{equation}
Let us consider one of the scalar products, denoted by $S(n_B,n_F)_{p,q}$, and
obtained from the scalar product of two basis vectors, first containing $C(n_p, n_F, p)$, second containing $C^{\dagger}(n_q, n_F, q)$.
We have
\begin{eqnarray}
S(n_B,n_F)_{p,q} 
&=& \langle \big\{n_{B} - n_p, 0\big\}| \big\{ C(n_p, n_{F}, p), C^{\dagger}(n_q, n_{F},q) \big\} | \big\{n_{B} - n_q, 0\big\}\rangle. \nonumber  \\
\label{eq.scalar product fermionic relation}
\end{eqnarray}
The anticommutator $\big\{ C(n_p, n_{F}, p), C^{\dagger}(n_q, n_{F}, q) \big\}$ is
a normally ordered operator containing only bosonic creation and
annihilation operators. If it is not, one can always bring it to such form\footnote{Such an anticommutator is equal to a
sum of operators involving only bosonic creation and annihilation operators. In general not all of these operators
will be normally ordered operators. However, any such operators can be brought
to a normally ordered form by appropriately ordering the creation and annihilation operators, which they are composed of,
using the commutation rules eqs.\eqref{eq:boz komutatory}. Such ordering
will produce additional operators which have to be taken into account and which can also be brought to a normally ordered.
Thus, any anticommutator can always be written in a normally ordered form.\label{footnote. normal ordering}}.
The matrix elements of the operator $C^{\dagger}(n_q, n_{F}, q) C(n_p, n_{F}, p)$ vanish since there are fermionic
annihilation operators acting on the Fock vacuum. Thus, by anticommuting
all fermionic operators in one step, we can express $S(n_{B}, n_{F})$ in terms of
matrix elements of operators between states from the bosonic sector exclusively.\\

Summarizing, the procedure of calculating $S(n_{B}, n_{F})$ consists of three steps. First, we pull out
one bosonic (fermionic) brick from the left and right states. There is in general $N-1$ ( $d(n_F)$ ) ways to do this.
Second, we commute (anticommute) these two bricks,
and replace the commutator (anticommutator) by a
normally ordered operator. Third, we evaluate the matrix elements of
this operator between states with a lower number of quanta (see section \ref{sec. Matrix elements of gauge invariant operators}). \\

However, before we move to the description of the evaluation of matrix elements of operators, we have to tackle the problem
of orthonormalization of the basis states and of the evaluation of the commutators of composite bricks. We do this in the following two subsections.


\subsubsection{Linear independence and orthonormalization}
\label{sec. Linear independence and orthonormalization}
The recursive approach produces a set of Fock states in which some states may be contained in several copies.
Therefore, one has to implement
a mechanism to remove such redundancy. Because of the recursive structure,
such mechanism will have to deal only with states with
a given number of bosonic and fermionic quanta. Hence, its computational effort is small compared to what would it
be should the whole Fock basis be considered.
Moreover, the remaining states must be orthonormalized.
We use a procedure which realizes these two tasks in one step. It is done by a numerical diagonalization
of the matrix of scalar products $S(n_B, n_F)$. Subsequently,
eigenvectors with corresponding nonzero eigenvalues are retained.
Since the Fock states are not normalized, the normalized eigenvectors have to be multiplied by the inverse of the square root of their corresponding
eigenvalues.
We group such vectors in a matrix denoted by $R(n_{B}, n_{F})$. Note that $R(n_{B}, n_{F})$ is not to a square matrix in general.
Then, we can write
\begin{equation}
R^T(n_{B}, n_{F}) S(n_B, n_F) R(n_{B}, n_{F}) = 1_{D(n_B,n_F) \times D(n_B, n_F)}, \nonumber
\end{equation}
where $1_{D(n_B,n_F) \times D(n_B, n_F)}$ is the unity matrix which rank is equal to the size of the subspace of the Hilbert space with $n_B$ bosonic and $n_F$ fermionic quanta.


\subsubsection{Automatic evaluation of commutators and anticommutators}
\label{sec. Automatic evaluation of commutators and anticommutators}

We can conclude from eqs. \eqref{eq.scalar product bosonic relation} and \eqref{eq.scalar product fermionic relation} that,
in order to calculate a scalar product or a matrix element of some operator,
the set of commutators and anticommutators of all composite bricks must be supplied. These commutators and anticommutators must
be brought to a gauge invariant, normally ordered form and should be maximally reduced using the Cayley-Hamilton theorem.
In general, there will appear some new single trace operators containing both creation and annihilation bosonic operators.
Their commutators and anticommutators with all composite bricks should be evaluated as well and supplied to the algorithm.
The number of such relations to be calculated grows rapidly, both, with increasing $N$, and increasing fermionic occupation number.
Already, for the $SU(3)$ gauge group
one needs about a thousand of (anti)commutators. Therefore, a computer program was written to evaluate them.
It uses standard 
relations \eqref{eq:boz komutatory}, \eqref{eq:ferm komutatory} and \eqref{eq:mixed komutatory}
between the bosonic and fermionic creation and annihilation operators to move them among and within traces.
As an example Eqs.\eqref{eq. sample commutator} present one of the commutators needed for the calculations,
\begin{align}
[ (a^{\dagger}a^{\dagger}aa^{\dagger}a^{\dagger}aa),(a^{\dagger}a^{\dagger}) ] &= 1.5555 (a^{\dagger}a^{\dagger}a^{\dagger}) + 1.3333 (a^{\dagger}a^{\dagger})(a^{\dagger}a^{\dagger}a) \nonumber \\
&+ (a^{\dagger}a^{\dagger}a^{\dagger})(a^{\dagger}a^{\dagger}aa) + 1.1111(a^{\dagger}a^{\dagger}a^{\dagger})(a^{\dagger}a) \nonumber \\
&+ 0.1666(a^{\dagger}a^{\dagger})(a^{\dagger}a^{\dagger}a^{\dagger})(aa) + (a^{\dagger}a^{\dagger})(a^{\dagger}a^{\dagger}aa^{\dagger}a) \nonumber \\
&+ 0.25(a^{\dagger}a^{\dagger})(a^{\dagger}a^{\dagger})(a^{\dagger}aa)
\label{eq. sample commutator}
\end{align}

We have found that it is more efficient to work with
composite bricks than elementary ones.
For example, if we have treated the operator
$(a^{\dagger}a^{\dagger}f^{\dagger})(a^{\dagger}f^{\dagger})(a^{\dagger}f^{\dagger}f^{\dagger})$,
which is one of the composite bricks in the $n_F=4$ sector for the $SU(3)$ group,
as a product of three operators, we would relate
the desired matrix element with matrix elements from the $n_F=3$ sector, which, in turn, would be related
to some matrix elements in the $n_F=2$ sector, and so on. However, one could treat it as a single operator,
and jump to the bosonic sector in one step, considerably decreasing the number of commutation and anticommutation
relations needed for such computations. On the other hand, the more elementary bricks will be contained in a composite brick,
the more complicated the commutation and anticommutation
relations will be. Hence, the usefulness of using more complex composite bricks is a question of balancing between
the processor time consumed for evaluation of these relations, the
memory needed to store them and the processor time gained by using more complex (anti)commutators.

\subsection{Matrix elements of gauge invariant operators}
\label{sec. Matrix elements of gauge invariant operators}

In this section we describe the evaluation of matrix elements of operators, such as those appearing in the
right hand side of eqs. \eqref{eq.scalar product bosonic relation} and \eqref{eq.scalar product fermionic relation}.\\

Let us denote a generic operator by $O(n^{O}_{B}, n^{O}_{F})$.
The arguments of $O$, $n^{O}_{B}$ and $n^{O}_{F}$ have the following meaning
\begin{itemize}
\item $n^{O}_{B}$ is the difference between the number of bosonic creation
and annihilation operators contained in $O$,
\item $n^{O}_{F}$ is the difference between the number of
fermionic creation and annihilation operators in $O$.
\end{itemize}
In general, $n^{O}_{B}$ and $n^{O}_{F}$ can be any integers. Furthermore, we denote by \newline $O(n^{O}_{B}, n^{O}_{F})_{n_{B}, n_{F}}$ the matrix element
of $O$ between basis states containing $n_B$ bosonic and
$n_F$ fermionic quanta on the right hand side, and $n'_B = n_B + n^{O}_B$ bosonic and $n'_F = n_F + n^{O}_F$ fermionic
 quanta on the left hand side,
\begin{equation}
O(n^{O}_{B}, n^{O}_{F})_{n_{B}, n_{F}} \ = \ \langle \big\{n'_{B}, n'_{F} \big\} |O(n^{O}_{B}, n^{O}_{F})| \big\{n_{B}, n_{F} \big\} \rangle.
\end{equation}
$O(n^{O}_{B}, n^{O}_{F})_{n_{B}, n_{F}}$ is a matrix of sizes $D(n_B, n_F) \times D(n'_B, n'_F)$,
where the numbers $D_{n, m}$ denote the multiplicity of basis states with $n$ bosonic and $m$ fermionic quanta,
as it was introduced in section \ref{sec. Character method}.\\

We first deal with some 'boundary' situations, and then we consider the generic case.

\subsubsection{Boundary cases}
\label{sec. Boundary cases}

Let $\#(x)_O$ denote the number of occurrences of the operator $x$ in the operator $O$.
The following observations can be exploited to simplify the computations:
\begin{itemize}
\item if a matrix element of an operator for which $\#(f^{\dagger})_O > \#(f)_O$
is to be calculated then it is more convenient to evaluate its complex conjugate. Similarly,
if we have to compute a matrix element of an operator for which $\#(f^{\dagger})_O = \#(f)_O$, but $\#(a^{\dagger})_O > \#(a)_O$
we should rather compute its complex conjugate.

\item the matrix element of an operator $O$ which has fermionic or bosonic annihilation operators acting on the Fock vacuum vanishes.

\item the matrix element of the bosonic elementary brick between states from the bosonic sector can be read off
from the appropriate part of the matrix of scalar products.

\item the matrix element of an operator which is a product of two trace operators can be calculated by inserting an identity operator
between them, evaluating their matrix elements
separately, and eventually multiplying  and summing the partial results.
\end{itemize}

\subsubsection{Generic case}
\label{sec. Generic case}

In the generic case, we can assume that $O$ is normally ordered (see footnote \ref{footnote. normal ordering} on page \pageref{footnote. normal ordering}).
If it is composed exclusively of creation operators, one can express it in terms of bosonic elementary
bricks and use the appropriate boundary case, described above.

The strategy to evaluate a matrix element of $O$ is
to drag $O$ over the operators constituting the right hand side state so that it annihilates the Fock vacuum.
We start by pulling it through the composite fermionic brick,
\begin{eqnarray}
\big( O(n^{O}_{B}, n^{O}_{F})_{n_{B}, n_{F}} \big)_{.,p}
&=& \langle \big\{n'_{B}, n'_{F}\big\} |O(n^{O}_{B}, n^{O}_{F})C^{\dagger}(n_p,n_F,p)| \big\{n_{B}, n_{F}\big\} \rangle \nonumber \\
&=& \langle \big\{n'_{B}, n'_{F}\big\}| \big[O(n^{O}_{B}, n^{O}_{F}), C^{\dagger}(n_p,n_{F}, p)\big] | \big\{n_{B}-n_p,0\big\}\rangle \nonumber \\
&+& \langle \big\{n'_{B}, n'_{F}\big\}| C^{\dagger}(n_p,n_{F}, p) O(n^{O}_{B}, n^{O}_{F}) | \big\{n_{B}-n_p,0\big\}\rangle. \nonumber \\
\label{eq.a nie wiem}
\end{eqnarray}
In order to move further, we substitute the relation for the (anti)commutator of $O(n^{O}_{B}, n^{O}_{F})$ and $C^{\dagger}(p,n_{F}, \alpha)$.
For each operator appearing in this relation we evaluate its matrix element, first checking whether conditions for any of the special cases are met.
This task should be easier, since these matrix elements must be evaluated between states with smaller
number of bosonic and fermionic quanta.
The second term in eq. \eqref{eq.a nie wiem},
can be calculated by
inserting an identity operator between the operators $C^{\dagger}(n_p,n_{F}, p)$ and $O(n^{O}_{B}, n^{O}_{F})$. Again, this computation
should be easier. On one hand, the
matrix element of the elementary brick $C^{\dagger}(n_p,n_{F}, p)$ should be known from the matrix of scalar products.
On the other hand, the matrix element of $O(n^{O}_{B}, n^{O}_{F})$ involves states with smaller
number of bosonic and fermionic quanta.

The purely bosonic case can be treated analogously. We have
\begin{align}
\big(O(n^{O}_{B},0)_{n_B,0}\big)_{.,p} &= \langle \big\{n'_{B},0\big\}| \big[O(n^{O}_{B},0), C(p, 0)\big] | \big\{n_{B}-p,0\big\}\rangle \nonumber \\
&+ \langle \big\{n'_{B},0 \big\}| C(p,0) O(n^{O}_{B},0) | \big\{n_{B}-p,0\big\}\rangle ,
\label{eq.a nie wiem2}
\end{align}

In principle, we can proceed with those relations until
$O$ hits the Fock vacuum. Since $O$ is a normally ordered operator, such matrix element
vanishes by definition.
Therefore, collecting all the intermediate results, we should be able to evaluate the desired matrix element.
Nevertheless, one has
to remember that states in $| \big\{n_{B},0 \big\} \rangle$ are in general not orthonormal, so one has to implement into relations
\eqref{eq.a nie wiem} and \eqref{eq.a nie wiem2} the orthonormalization procedure.

\subsubsection{Recurrence relations}
\label{sec. Recurrence relations}

Since the recursively constructed set of basis states can contain degenerate states,
the matrices of operators, calculated as described above,
will also contain such redundant matrix elements. In order to get rid of them one has to use the $R(n_{B}, n_{F})$ matrix.
Incorporating this matrix in the relations for the matrix element of any operator, enables us
to formulate the complete and correct recurrence relations.
We have two recurrence relations:
\begin{itemize}
\item the expression of the matrix element in the fermionic sectors in terms of matrix elements in the bosonic sector,
\end{itemize}
\begin{align}
\langle \big\{n'_{B}, n'_{F}\big\} &|O(n^{O}_{B}, n^{O}_{F})| \big\{n_{B}, n_{F}\big\} \rangle  = \nonumber \\
&\Big( \langle \big\{n'_{B}, n'_{F}\big\}| \big[O(n^{O}_{B}, n^{O}_{F}), C^{\dagger}(n_p,n_{F}, p)\big] | \big\{n_{B}-n_p,0 \big\}\rangle \nonumber \\
&+ \langle \big\{n'_{B}, n'_{F}\big\}| C^{\dagger}(n_p,n_{F}, p) O(n^{O}_{B}, n^{O}_{F}) | \big\{n_{B}-n_p,0\big\}\rangle \Big) \cdot R (n_{B}, n_{F}),
\label{eq. correct fermionic recursion relation}
\end{align}
and
\begin{itemize}
\item the expression of the matrix element in the bosonic sector with bigger number of bosonic quanta
in terms of matrix elements with smaller number of bosonic quanta,
\end{itemize}
\begin{align}
\langle \big\{n'_{B},0 \big\} &|O(n^{O}_{B},0)| \big\{n_{B},0\big\} \rangle  =
\Big( \langle \big\{n'_{B},0 \big\}| \big[O(n^{O}_{B},0), C^{\dagger}(p, 0)\big] | \big\{ n_{B}-p,0 \big\} \rangle \nonumber \\
&+ \langle \big\{ n'_{B},0 \big\} | C^{\dagger}(p,0) O(n^{O}_{B},0) | \big\{ n_{B}-p,0 \big\} \rangle \Big) \cdot R (n_{B},0).
\label{eq. correct bosonic recursion relation}
\end{align}
Thus, we have expressed the desired matrix element in terms of matrix elements of operators between states will lower
number of quanta, which should have been already evaluated during some previous calculations.

With these recurrence relations
the presentation of the whole algorithm is complete.
One can use
them to evaluate the matrix of scalar products in the sectors which have not been considered
so far in the calculations. This done, the matrix elements of operators, needed for the calculations in sectors
with yet bigger number of quanta, can be computed. In this way one can proceed until the cut-off $N_{cut}$ is reached.

Note that the above algorithm is very universal. In principle it can be used to systems defined in space of any dimensionality.
Particularly, if the Hamiltonian is invariant under a $SO(d)$ symmetry
our method can be generalized in order to calculate the spectra in the channels with given angular momentum \cite{wosiek5}.
Moreover, it is applicable to systems with bosonic and fermionic
polynomial interactions as well as to systems with discrete or continuum spectrum.

In the next section we describe the results for supersymmetric model with discrete spectrum obtained with our approach.

\section{Applications}
\label{sec. Applications}


As an application of the above algorithm we present results for a supersymmetric system given by the
Hamiltonian constructed in \ref{sec. Supersymmetric Yang-Mills Quantum Mechanics},
\begin{equation}
H = \textrm{tr} \ p^2 + g^2 \Big( \textrm{tr} \ x^4 - \frac{1}{3} (\textrm{tr} \ x^2)^2 \Big) + 2g \ \textrm{tr}(x [ f^{\dagger}, f]),
\label{eq. hamiltonian dyskretny}
\end{equation}
where the gauge symmetry group is chosen to be the $SU(3)$ group.

The main motivation for studying this Hamiltonian is that, being a
system with a discrete spectrum, it is an good test-ground for our numerical method. Moreover,
the bosonic part of its potential is similar to the potential of SYMQM with the tensor $f_{abc}$ replaced by $d_{abc}$ \cite{wosiek5}.
Therefore, one hopes that some analytic approaches based on the numerical results from such simple model can be tested
for a future application to the more dimensional SYMQM systems. On the other hand, eq.\eqref{eq. hamiltonian dyskretny} is interesting
by itself. Being an supersymmetric anharmonic oscillator it contains a nontrivial fermionic interaction.

In the following we start by analyzing the convergence of the eigenvalues, then present the spectra calculated numerically.
Subsequently, we briefly describe their symmetries, such as supersymmetry and scaling
symmetry. Eventually, we calculate
numerically the Witten index for this model.

\subsection{Numerical spectra}

\begin{figure}[!htb]
\begin{center}
\input{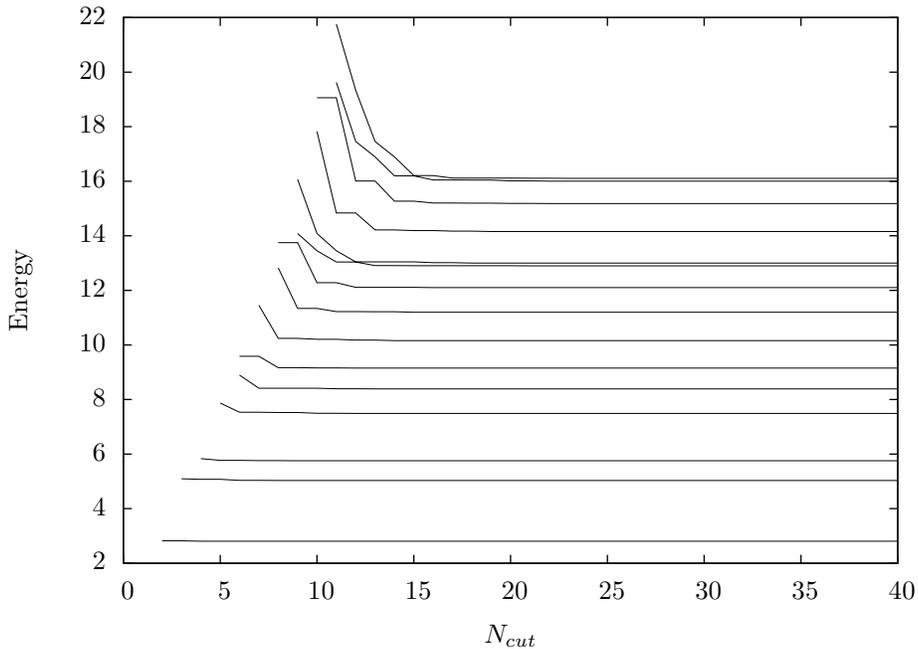}
\caption{Convergence of the 15 lowest eigenenergies in the bosonic sector with increasing cut-off. An exponential-like convergence
is seen, which is a characteristic feature of bound states. \label{fig.d.N_dep_f0}}
\end{center}
\end{figure}
\begin{figure}[!htb]
\begin{center}
\includegraphics[width=0.33\textwidth,angle=270]{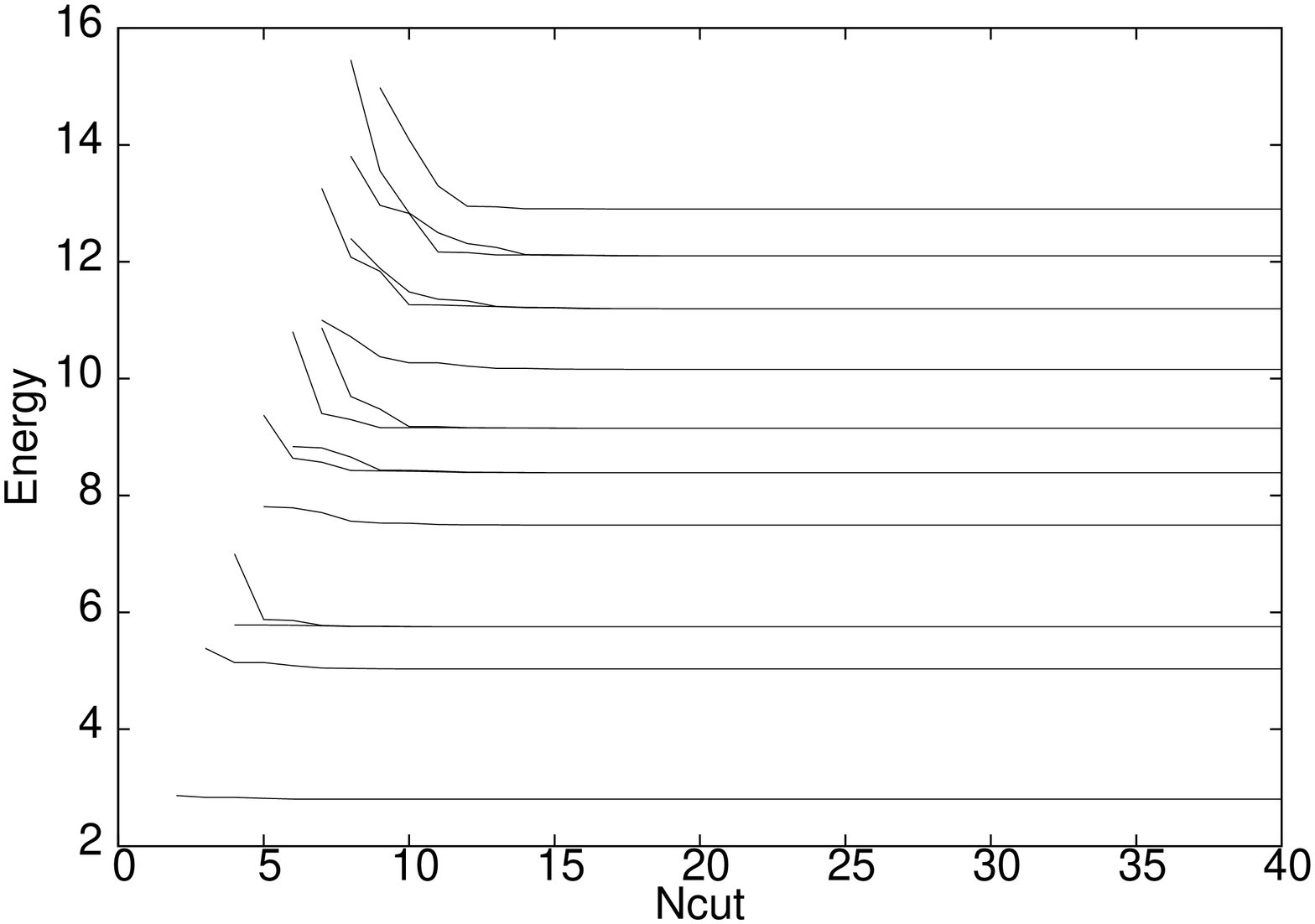}
\includegraphics[width=0.33\textwidth,angle=270]{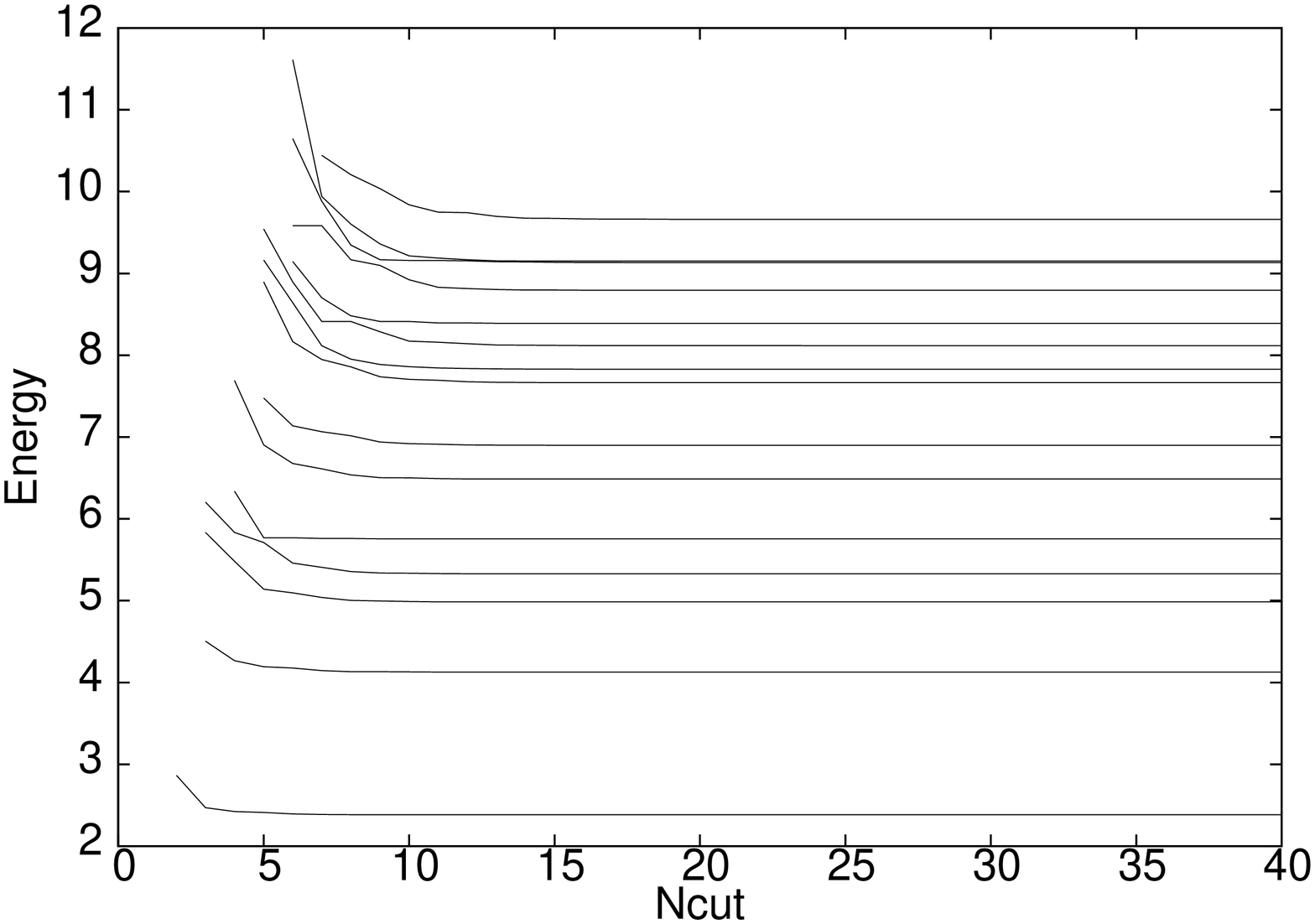}
\includegraphics[width=0.33\textwidth,angle=270]{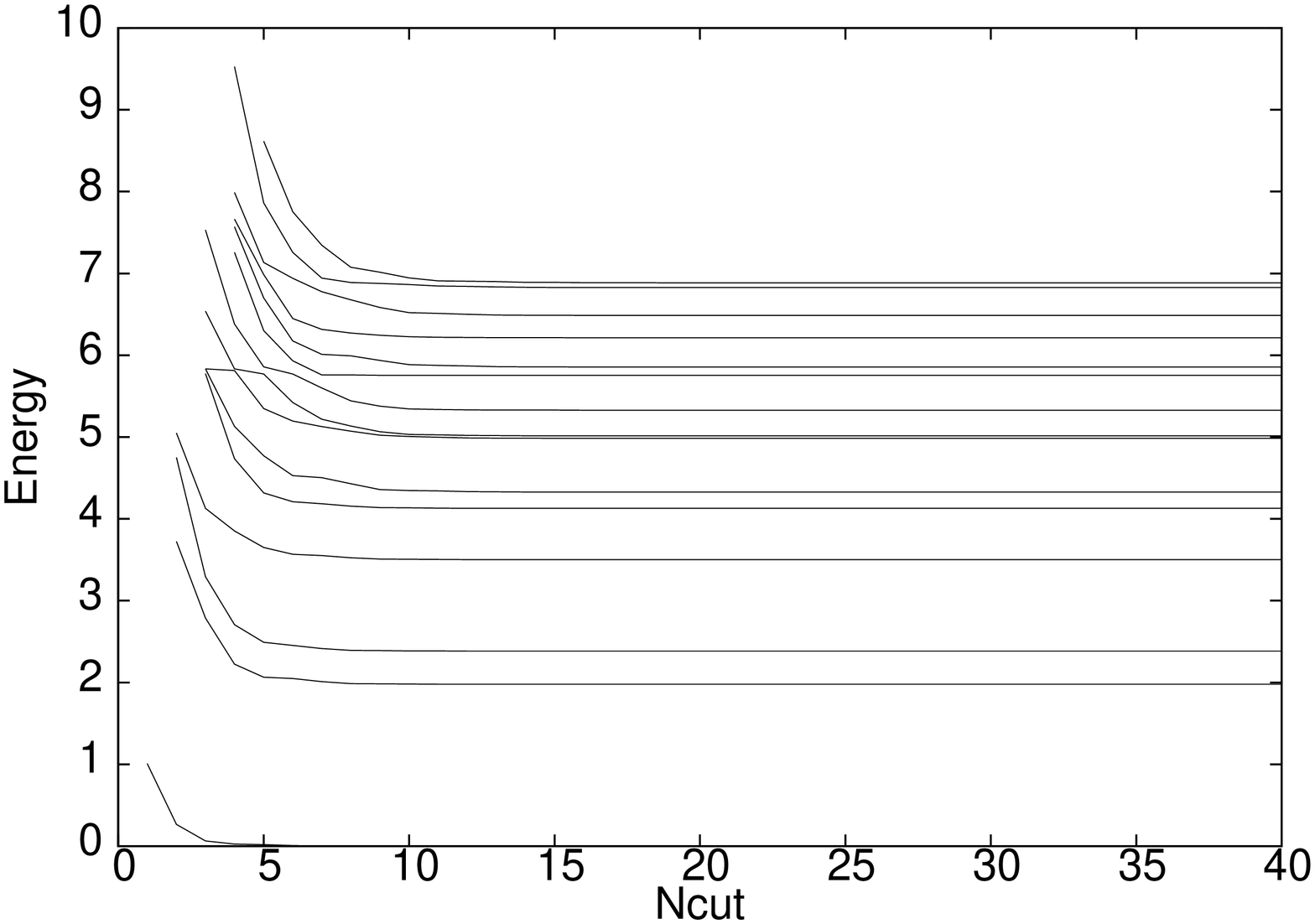}
\includegraphics[width=0.33\textwidth,angle=270]{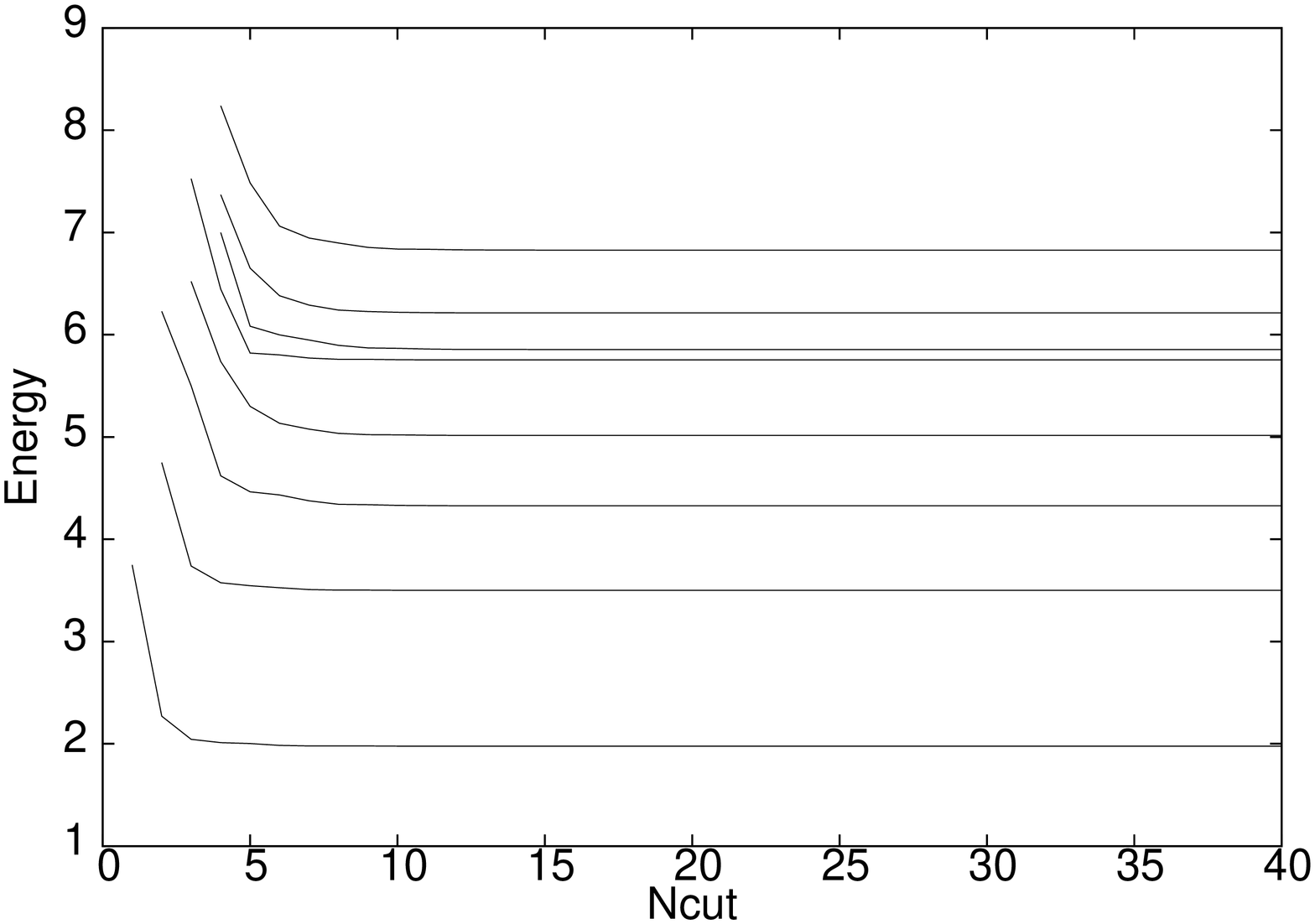}
\caption{Convergence of the few lowest eigenenergies in the $n_F=1,...,4$ sectors with increasing cut-off. \label{fig.d.N_dep_f1}}
\end{center}
\end{figure}

One of the advantages of the cut Fock space approach as a numerical method is that it enables one to
judge on the reliability of the results and estimate their errors. To this end, the convergence of eigenenergies or eigentates
with increasing cut-off must be investigated. Table \ref{tab. convergence} contains the energies of the lowest eigenstate in fermionic sector with $n_F=0,\dots,4$
obtained for different $N_{cut}$.
\begin{table}[h!]
\begin{center}
\begin{tabular}{|c|l|l|l|}
\hline
$N_{cut}$ & $n_F=0$ & $n_F=1$ & $n_F=2$ \\
\hline
\hline
1 &   2.833333333333333 & 3.750000000000124 & 3.75000000000008 \\
5 &   2.805137759654418 & 2.817654396966426 & 2.41010649311797 \\
10 &  2.804878933491876 & 2.804943385906189 & 2.38393952020263 \\
15 &  2.804877899477374 & 2.804878578502977 & 2.38379874405844 \\
20 &  2.804877857980324 & 2.804877869314702 & 2.38379576457689 \\
25 &  2.804877857812559 & 2.804877857890121 & 2.38379573799721 \\
30 &  2.804877857802534 & 2.804877857804384 & 2.38379573773261 \\
35 &  2.804877857802529 & 2.804877857803605 & 2.38379573772474 \\
40 &  2.804877857802507 & 2.804877857803596 & 2.38379573772458 \\
\hline
$N_{cut}$ & $n_F=3$ & $n_F=4$ & \\
\hline
\hline
1 &   1.009109012532963 & 3.750000000000082 & \\
5 &   0.017808308382480 & 2.001903558629864 & \\
10 &  0.000102896003680 & 1.978068350121234 & \\
15 &  0.000002580129746 & 1.977963562445323 & \\
20 &  0.000000018188518 & 1.977960963230507 & \\
25 &  0.000000000380532 & 1.977960939859982 & \\
30 &  0.000000000013940 & 1.977960939644200 & \\
35 &  0.000000000004579 & 1.977960939638051 & \\
40 &  0.000000000004250 & 1.977960939637698 & \\
\hline
\end{tabular}
\end{center}
\caption{The dependence on the cut-off $N_{cut}$ of the lowest eigenenergy  in several fermionic sectors. Note the exact degeneracy of the levels
in the $n_F=0$ and $n_F=1$ sectors due to supersymmetry and the appearance of the supersymmetric vacuum in the $n_F=3$ sector. \label{tab. convergence}}
\end{table}
The results in sectors with higher fermionic occupation number are related to the ones presented in the table through the particle-hole symmetry.
A convergence to more than 10 digits is achieved with $N_{cut} = 40$, which corresponds to a Fock basis of
about 150 states in the bosonic sector and about 1300 states in the $n_F=4$ sector.
Therefore, one can safely use the cut-off $N_{cut}=40$ in order to evaluate the lowest eigenenergies.
The uncertainties of those eigenenergies,
defined as the difference of the outcomes for consecutive cut-offs, are negligible.
Figure \ref{fig.d.N_dep_f0} and figures \ref{fig.d.N_dep_f1} show
the dependence of 15 lowest eigenenergies on the cut-off in different fermionic sectors.
All the figures were made for $g=1.0$. We can clearly see a fast,
exponential-like, convergence, which was shown to be a characteristic feature of bound states \cite{maciek+wosiek}\cite{maciek3}.
Hence, one can conclude that indeed all the spectra are discrete. This should be contrasted with the results for the SYMQM systems. The latter
have potentials with flat directions which induce continuum spectra in some sectors \cite{wosiek1}\cite{wosiek5}\cite{praca_magisterska}.
Table \ref{tab.susy_pairing} contains the eigenenergies of lowest states in all 9 fermionic sectors calculated with $g=1.0$ and a cut-off
$N_{cut} = 40$. We comment these results in more details in the following subsection.
\begin{table}[!htb]
\begin{center}
\footnotesize
\begin{tabular}{|c|c|c|c|c|c|c|c|c|}
\hline
$n_F=0$     &    $n_F=1$    &   $n_F=2$     &   $n_F=3$     &   $n_F=4$     &   $n_F=5$     &   $n_F=6$     &   $n_F=7$     & $n_F=8$       \\
\hline
\hline
        &           &           & 0         &           & 0         &           &           &           \\
        &           &           & 1.97796   & 1.97796$^2$   & 1.97796   &           &           &           \\
        &           & 2.3838    & 2.3838    &           & 2.3838    &   2.3838  &           &           \\
2.80488 & 2.80488   &           &           &           &           &           &   2.80488 &  2.80488  \\
        &           &           & 3.50147   & 3.50147$^2$   & 3.50147   &           &           &           \\
        &           & 4.12774   & 4.12774   &           & 4.12774   &   4.12774 &           &           \\
        &           &           &  4.32778 & 4.32778$^2$   & 4.32778   &           &           &           \\
        &           & 4.98444      & 4.98444     &            & 4.98444      &   4.98444    &           &           \\
        &           &           &  5.01572   & 5.01572$^2$   & 5.01572   &           &           &           \\
5.02988 & 5.02988   &           &           &           &           &           &   5.02988 &  5.02988  \\
        &           & 5.32891   & 5.32891   &           & 5.32891   &   5.32891 &           &           \\
5.75469 & 5.75469   &           &           &           &           &           &   5.75469 &  5.75469  \\
     & 5.75469   &   5.75469    &           &           &           &   5.75469   &   5.75469 &    \\
        &           &           & 5.75469 & 5.75469$^2$   & 5.75469   &           &           &           \\
        &           &           &  5.85572   & 5.85572$^2$   & 5.85572   &           &           &           \\
        &           &           &   6.21387   & 6.21387$^2$   & 6.21387   &           &           &           \\
        &           & 6.48704   & 6.48704   &           & 6.48704   &   6.48704 &           &           \\
        &           &           &  6.82769    & 6.82769$^2$   & 6.82769   &           &           &           \\
        &           &           &  6.88508   & 6.88508$^2$   & 6.88508   &           &           &           \\
        &           & 6.89869   & 6.89869  &           & 6.89869   &   6.89869 &           &           \\
7.49223 & 7.49223   &           &           &           &           &           &  7.49223 &  7.49223  \\
        &           & 7.66421   & 7.66421   &           & 7.66421   &   7.66421 &           &           \\
        &           &           &  7.75149  & 7.75149$^2$   & 7.75149   &           &           &           \\
        &           &           & 7.81189    & 7.81189$^2$   & 7.81189   &           &           &           \\
        &           & 7.82955   & 7.82955   &           & 7.82955   &   7.82955 &           &           \\
        &           & 8.11747   & 8.11747   &           & 8.11747   &   8.11747 &           &           \\
8.39012 & 8.39012   &           &           &           &           &           &   8.39012 &  8.39012  \\
        & 8.39012   & 8.39012   &   &           &    &   8.39012 & 8.39012          &           \\
        &           &           &  8.39012  & 8.39012$^2$   & 8.39012   &           &           &\\
\hline
\end{tabular}
\caption{Eigenenergies of few lowest eigenstates with $g=1.0$. The states in the $n_F=4$ sector are double degenerate which follows from the particle-hole symmetry. Hence the $^2$ notation.
An exact, supersymmetric pairing can be observed among
states in adjacent fermionic sectors.
\label{tab.susy_pairing}}
\end{center}
\end{table}
\normalsize

\subsection{Symmetries}

\subsubsection{Supersymmetry}

Supersymmetry can be seen in table \ref{tab.susy_pairing} as an exact degeneracy of the converged eigenenergies in neighboring fermionic
sectors. One notices a non-degenerate vacuum state in sector with $n_F=3$ and its
image through the particle-hole symmetry in the sector with $n_F=5$ signaling an unbroken supersymmetry. The number of
supersymmetric vacua was discussed in \cite{maciek4} in the case of free $SU(N)$ model. Accordingly, there are four zero-energy states
in sectors with $n_F=0,3,5,8$. Hence, two of these states disappear when the interaction is turned on.
The particle-hole symmetry is responsible
for a double degeneracy of eigenenergies in the sector with $n_F=4$. This sector must contain as many states which are parts of supermultiplets
formed with states from the sector with $n_F=3$ as there are states which form supermultiplets with states from the sector with $n_F=5$. Moreover,
both sets of supermultiplets must have the same spectra.
This is a specific feature of models with $SU(N)$ gauge group with $N$ odd.

Table \ref{tab. convergence} reveals also signatures of supersymmetry.
One one hand, the lowest states in sectors with $n_F=0$ and $n_F=1$ are exactly degenerate. On the other hand,
the presence of a non-degenerate supersymmetric vacuum state in the sector with $n_F=3$ and in the sector with $n_F=5$
is a proof that the supersymmetry is unbroken in this model.

\subsubsection{Scaling symmetry}

The scaling property of the quantum anharmonic oscillator was first noted by Symanzik, and elaborated by Simon \cite{simon}. If we
consider the transformations
\begin{equation}
x \rightarrow \lambda x, \quad p \rightarrow \frac{1}{\lambda} p, \qquad f \rightarrow f, \quad f^{\dagger} \rightarrow f^{\dagger},
\label{eq. transformation}
\end{equation}
then the Hamiltonian is rescaled as
\begin{equation}
H \rightarrow H = \frac{1}{\lambda^2} \Bigg( \textrm{tr} \ p^2 + g^2  \lambda^6 \Big( \textrm{tr} \ x^4 - \frac{1}{N} (\textrm{tr} \ x^2)^2 \Big) + 2 g \ \lambda^3 \ \textrm{tr} (x [ f^{\dagger}, f]) \Bigg).
\end{equation}
Setting $\lambda = g^{-\frac{1}{3}}$ we obtain the following identity,
\begin{equation}
H(g) \rightarrow g^{\frac{2}{3}} H(1).
\label{eq. hamiltonian scaling}
\end{equation}
Since, the transformations eq.(\ref{eq. transformation}) can be unitarily implemented, both Hamiltonians in eq.(\ref{eq. hamiltonian scaling})
have identical eigenvalues. Therefore, it is sufficient to calculate the spectrum at $g=1.0$.

Figure \ref{fig.d.g_dep} shows the dependence of six lowest eigenenergies from the bosonic sector on the coupling constant. The numerical results
are compared with
the prediction of eq.\eqref{eq. hamiltonian scaling}.
For large $g$ the agreement of both should be noted, whereas
the discrepancies for small $g$ are due to finite cut-off effects as discussed below.

\subsection{Critical slowing down}

\begin{figure}[h!]
\begin{center}
\input{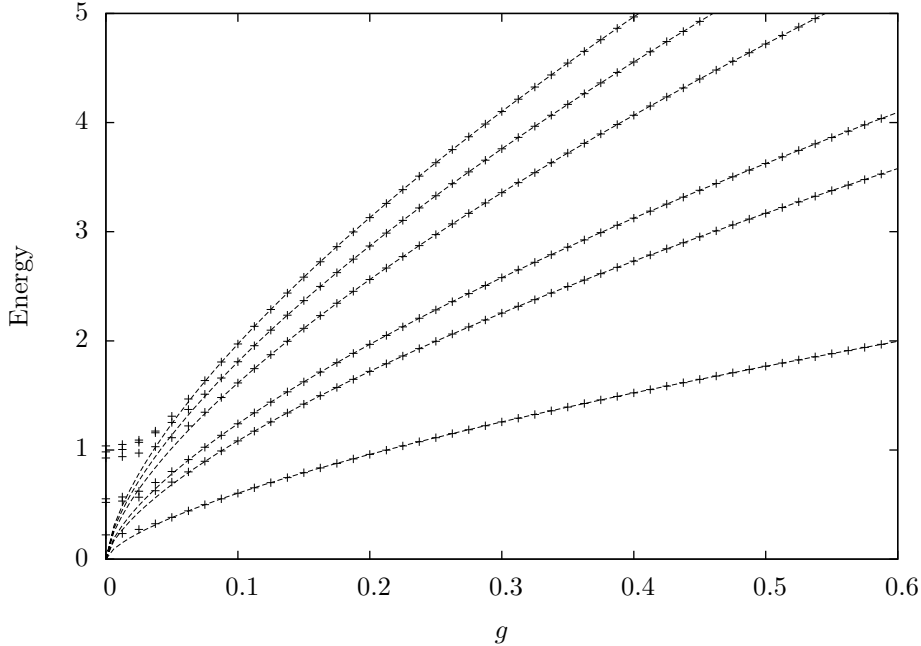}
\caption{Dependence of the first six eigenenergies on the coupling constant $g$ in the bosonic sector. Crosses
denote the numerical results obtained with cut-off $N_{cut}=40$, whereas the dashed lines correspond
to the prediction of eq. \eqref{eq. hamiltonian scaling} with $H(1)$ evaluated numerically. The finite cut-off effects can be seen for small values of
the coupling constant.\label{fig.d.g_dep}}
\end{center}
\end{figure}

A critical slowing down can be observed on figure \ref{fig.d.g_dep} in the vicinity of $g=0$.
For a small value of the coupling constant a much higher cut-off is needed in order to obtain converged results.
It is because at $g=0$ the spectrum is free and the
eigenenergies corresponding to nonlocalized states calculated by our algorithm do not converge.
Rather, they fall off to zero with increasing cut-off in a power like manner, a behavior resulting
from approximating a plane wave by a finite set of localized harmonic oscillator eigenstates.
Hence, the eigenenergies calculated for small coupling constant at small cutoff cannot follow the curve eq.\eqref{eq. hamiltonian scaling}
of exactly converged energies.
Nevertheless, with increasing coupling constant the eigenenergies of bound states converge more and more rapidly.
Those energies that have already converged with the cut-off agree with the analytic prediction of eq.\eqref{eq. hamiltonian scaling}.
The complete discussion of such critical behavior is out of scope of the present work
and will be carried out elsewhere.

\subsection{Perturbative expansion}

\begin{figure}[h!]
\begin{center}
\input{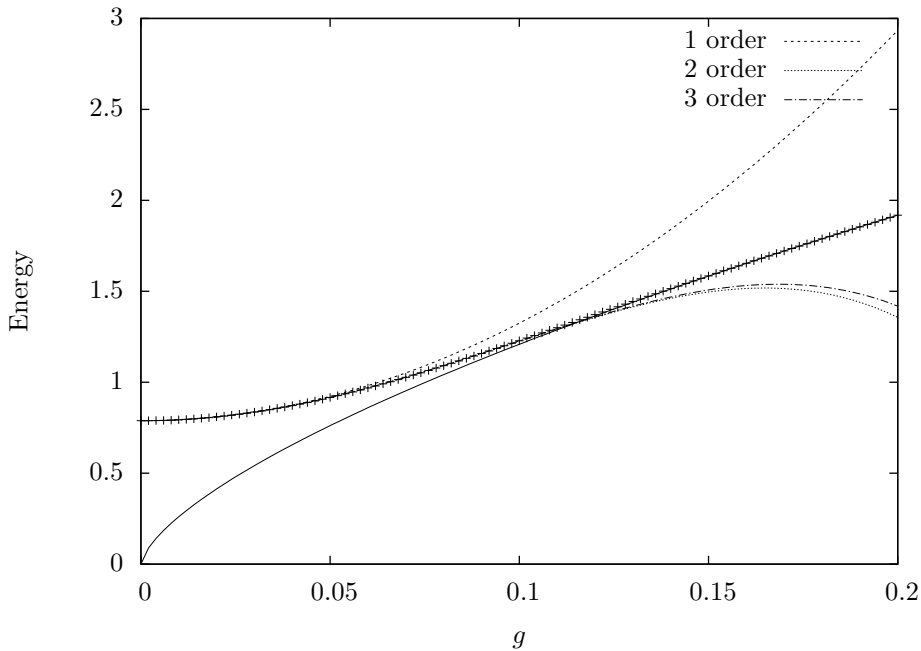}
\caption{Comparison of the dependence of the energy of the lowest bosonic eigenstate on the coupling constant $g$.
Numerical data (crosses) are plot together with
the prediction of eq.\eqref{eq. hamiltonian scaling} (solid line), and the perturbative expansion eq.\eqref{eq. perturbative}(dashed lines)
 \label{fig. perturbation}}
\end{center}
\end{figure}

In a cut Fock space the states corresponding to the free states in the continuum are normalizable.
Moreover, their exact analytic form can be obtained for a finite cut-off \cite{korcyl1.5}\cite{korcyl3}.
Therefore, \emph{at fixed finite cut-off $N_{cut}$} one can
perform a perturbative expansion in the coupling constant $g$.
Using well-known formulae for the perturbative corrections
to the energy \cite{qm1} one can obtain the approximate dependence of the eigenenergies on the coupling constant.
We present here the results
for the lowest eigenstate from the bosonic sector calculated for the cut-off $N_{cut}=20$,
\begin{align}
E &= E_0 + g^2 V_{E_0,E_0} + g^4 \sum_{E' \ne E_0} \frac{\big|V_{E_0, E'} \big|^2 }{E_0 - E'} +\nonumber \\
& + g^6 \Big( \sum_{\substack{E' \ne E_0 \\ E'' \ne E_0}} \frac{V_{E_0, E'} V_{E', E''} V_{E'', E_0}}{(E_0 - E')(E_0 - E'')} - V_{E_0, E_0} \sum_{E' \ne E_0} \frac{ \big|V_{E_0, E'} \big|^2}{(E_0 - E')^2} \Big) \nonumber \\
&= 0.788363 + 53.6563 g^2 - 986.556 g^4 + 977.818 g^6,
\label{eq. perturbative}
\end{align}
where $V_{E,E'}$ is the matrix element of the potential between states with energies $E$ and $E'$ respectively.
The comparison of this formula with the numerical results
and the prediction of eq.\eqref{eq. hamiltonian scaling} is presented on figure \ref{fig. perturbation}.
It is remarkable that the perturbative expansion attains the values of the coupling constant
where the numerical results have converged for cut-off $N_{cut}$ and agree with eq.\eqref{eq. hamiltonian scaling}.
This gives us hope that with an improved perturbative expansion the whole spectrum may be obtained
analytically. This issue is being investigated in more details.

\subsection{Witten's index}

\begin{figure}[h!]
\begin{center}
\input{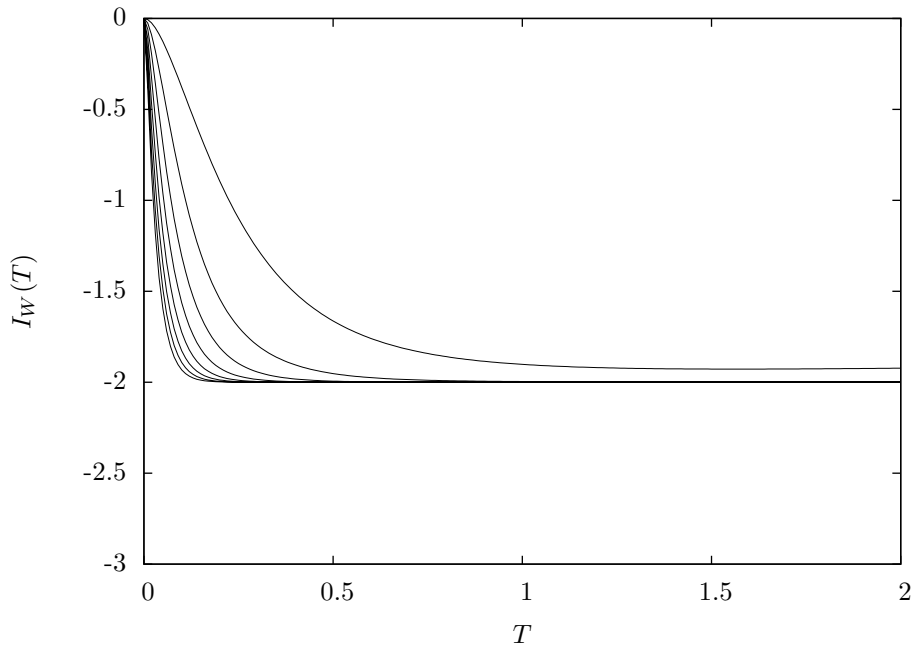}
\caption{Witten index. Different curves correspond to increasing cut-offs: 5, 10, 15, ... , 40. \label{fig.d.witten}}
\end{center}
\end{figure}

The Witten index is defined as \cite{witten1}\cite{witten2}
\begin{equation}
I_W(T) = \sum_{b \in \textrm{bosonic states}} e^{- E_b T} - \sum_{f \in \textrm{fermionic states}} e^{- E_f T}.
\label{eq. witten index}
\end{equation}
It is a commonly used quantity to study supersymmetry in quantum mechanics. In our set up, the sums over the bosonic and
fermionic states are finite due to the cut-off. For a given $N_{cut}$ one can just plug the eigenenergies obtained
from each fermionic sector into eq.\eqref{eq. witten index}. Our numerical results are shown on
figure \ref{fig.d.witten}, where the dependence of $I_W(T)$ on the euclidean time $T$ is presented.
One notices a rapid convergence of $I_W(T)$
to the value $-2$. This confirms the fact that the model has two supersymmetric vacua.

The collapse of the Witten index at $T=0$ to zero can be explained in the following way.
At $T=0$ the $I_W(T)$ is just the difference in the numbers of bosonic and fermionic states. However, the cut Fock basis
has an equal number of bosonic and fermionic states (see subsection \ref{sec. Symmetries of the $SU(N)$ SYMQM Fock basis}).
Thus, the value of the Witten index at $T=0$ is zero for any cut-off. This is no longer true at $T \ne 0$.
Especially, for $T \rightarrow \infty$, contributions from the states with nonzero energies cancel, and $I_W(T)$ counts
the numbers of supersymmetric vacua, equal $-2$ in the present case.

\section{Conclusions}
\label{sec. Conclusions}

    In this paper we have described in a very detailed manner a recursive algorithm for evaluation of matrix elements of any gauge
invariant operator in a Fock basis of Hilbert space. It can be applied to systems with any $SU(N)$ gauge group
and allows the evaluation of spectra in all fermionic sectors. We demonstrated the calculations on an example of anharmonic oscillator with
supersymmetric interactions for the $SU(3)$ group.

    We started by describing the idea of the numerical approach to quantum mechanics in the Hamiltonian formulation
    using the cut Fock space method. Then, we
    presented the construction of gauge-invariant Fock basis, and particularly we introduced the concepts of elementary
    bosonic bricks and composite fermionic bricks. Such approach provided us a systematic, recursive description of the Fock states
    with increasing number of quanta.
    We discussed the properties and symmetries of such basis. Next, we concentrated
    on the numerical algorithm. We described the calculation of the matrix of scalar products, emphasizing the main ideas of the recursion
    relations. Then, other parts of the algorithm were presented: the procedure which removes redundant basis vectors and
    orthonormalizes the remaining ones as well as the program which automatically calculates the commutators and anticommutators
    of given operators. Finally, the expressions for the evaluation of matrix elements of any operator were outlined, and eventually,
    the full recursion relations were presented.
    In the third part of this article we applied our algorithm to a supersymmetric system with $SU(3)$ gauge group and a discrete spectrum.
    We used it as a particularly well suited test-ground for our approach.
    We calculated the eigenenergies in all 9 fermionic sectors and discussed their symmetries.
    Eventually, we were also able to obtain the Witten index for this system.

    The main advantage of this algorithm is that it treats bosons and fermions on an equal footing, and thus, enables calculations
    in any fermionic sector of the Hilbert space. This should be contrasted with the sign problems
    encountered in lattice field theories. As a result an exact supersymmetric degeneracy can be obtained even for finite cut-off.
    Moreover, the approach can be applied to systems with discrete and continuous spectra
    as well as possessing any kind of gauge symmetry.
    Particularly, the supersymmetric anharmonic oscillator presented in this article with gauge groups with $N\ge3$\ is
    currently investigated. Similarly, the $D=2$, supersymmetric Yang-Mills quantum mechanics
    with several gauge groups, such as $SU(3)$, $SU(4)$ and $SU(5)$ are studied analytically and numerically \cite{korcyl1.5}\cite{korcyl3}.
    The flexibility of the algorithm enables also an generalization
    to higher dimensions, with the ultimate $D=10$, SYMQM case in mind. Results of the work is this direction are promising.

\section*{Acknowledgments}

The Author would like to thank for discussions with prof. J. Wosiek on the
subject of this paper and his careful reading of the manuscript.

\end{document}